\documentclass[manuscript]{acmart}
\usepackage{xcolor}
\usepackage{ulem}
\usepackage{multirow}
\usepackage{makecell}
\usepackage{multicol}
\usepackage{booktabs}
\usepackage{comment}
\usepackage{appendix}
\usepackage{cancel}
\usepackage{amsmath}
\usepackage{adjustbox}
\usepackage{tabularx}
\usepackage{booktabs}
\usepackage{siunitx}
\usepackage{mathtools}
\usepackage{graphicx}
\usepackage{placeins}
\AtBeginDocument{%
  }

\setcopyright{acmlicensed}
\copyrightyear{2025}
\acmYear{2025}
\acmDOI{XXXXXXX.XXXXXXX}

\acmJournal{TOCHI}

\begin{document}

\title{Designing Smarter Conversational Agents for Kids: Lessons from Cognitive Work and Means-Ends Analyses}

\author{Vanessa Figueiredo}
\orcid{0000-0001-9190-650X}
\affiliation{%
  \institution{University of Regina}
  \city{Regina}\state{Saskatchewan}\country{Canada}}
\email{vanessa-figueiredo@uregina.ca}

\renewcommand{\shortauthors}{Figueiredo}
\acmArticleType{Research}
\acmCodeLink{https://github.com/explorailab/child-ca-scaffolding-recipe}

\begin{abstract}
This paper presents two studies on how Brazilian children (ages 9–11) use conversational agents (CAs) for schoolwork, discovery, and entertainment, and how structured scaffolds can enhance these interactions. In Study 1, a seven‑week online investigation with 23 participants (children, parents, teachers) employed interviews, observations, and Cognitive Work Analysis to map children’s information‑processing flows, the role of more knowledgeable others, functional uses, contextual goals, and interaction patterns to inform conversation‑tree design. We identified three CA functions—School, Discovery, Entertainment—and derived “recipe” scaffolds mirroring parent–child support. In Study 2, we prompted GPT‑4o‑mini on 1,200 simulated child–CA exchanges, comparing conversation-tree recipes based on structured-prompting to an unstructured baseline. Quantitative evaluation of readability, question count/depth/diversity, and coherence revealed gains for the recipe approach. Building on these findings, we offer design recommendations: scaffolded conversation-trees, child‑dedicated profiles for personalized context, and caregiver‑curated content. Our contributions include the first CWA application with Brazilian children, an empirical framework of child–CA information flows, and an LLM‑scaffolding “recipe” (i.e., structured-prompting) for effective, scaffolded learning.
\end{abstract}




\maketitle

\section{Introduction}
Conversational agents (CAs) are designed to engage in communication exchanges that mimic human conversations \cite{sciuto_hey_2018, garg_last_2022}. These agents include digital assistants like Google Assistant, Siri, and Alexa, as well as chatbots and social robots. Although CAs are primarily designed to help users automate and manage tasks, children have also utilized CAs for academic purposes, such as information-seeking \cite{lovato_siri_2015, lovato_hey_2019}. However, most commercial CAs may not sufficiently support the development of children's independent problem-solving and critical thinking skills \cite{garg_conversational_2020}. These agents often provide ready-made solutions that limit a child's agency in academic activities, hindering children's academic and cognitive progress \cite{williams_is_2019, garg_conversational_2020}. While previous research has attempted to address these limitations by integrating knowledge scaffolding into the design of child- and education-oriented CAs \cite{alaimi_pedagogical_2020, han_teachers_2024}, there remains a limited understanding of how these approaches can be tailored to support children's information processing, concurrent activities (such as academic, knowledge discovery and entertainment), and collaborations with family members and peers.

Children's mental models tend to anthropomorphize these exchanges, attributing human-like qualities to CAs \cite{xu_are_2021}. However, CAs face several constraints, such as limitations in mutual understanding, building trust, and engaging in active listening—all fundamental aspects of human-human conversations \cite{hiniker_can_2021, lovato_hey_2019}. Therefore, human-CA interactions should not be entirely designed as human-human or human-computer (i.e., command-response) exchanges but should instead intersect elements from both communication types.

Previous research has incorporated knowledge scaffolding to enhance these communication exchanges and support children's problem-solving and critical thinking skills \cite{lee_dapie_2023, wang_informing_2022, rubegni_dont_2022}. These approaches draw on developmental and educational psychology theories, such as Vygotsky's Zone of Proximal Development (ZPD). ZPD posits that children need support from more knowledgeable others to understand new concepts \cite{vygotsky_mind_1978, wood_role_1976}. When applied to CA design, ZPD becomes knowledge scaffolding, where CAs pose questions to encourage children's agency and conceptual understanding \cite{alaimi_pedagogical_2020, xu_rosita_2023}. These scaffolding methods are focused solely on academic activities \cite{winkler_sara_2020, cheng_scientific_2024}, however, in reality, children may collaborate with family members and peers while interacting with CAs \cite{van_brummelen_what_2023}.

To address these gaps, we carried out two complementary studies.
\textbf{Study 1} was a seven‑week online investigation with 23 Brazilian participants (9–11 year‑old children, their parents, and teachers). Our goal was to uncover the interplay between system constraints, children’s information processing, activities, contexts, and personal characteristics by answering the following questions:

\begin{itemize}
    \item RQ1: How do children structure their information-processing flows when interacting with conversational agents (CAs)? 
    \item RQ2: What roles do more knowledgeable others  play in children’s CA-supported problem solving?
    \item RQ3: For what functional purposes do children use CAs, and what workarounds do they develop when CAs fall short?
    \item RQ4: What goals, constraints, and priorities influence children’s CA interactions in home contexts?
    \item RQ5: How can patterns in children’s CA interactions be translated into conversation-tree structures that better scaffold prompt formulation, problem solving, and engagement?
\end{itemize}

Through interviews, observations, critical‑incident sessions, and participant validation (\,\(\approx 23\)\ hours of video data), we applied Cognitive Work Analysis (CWA) \cite{rasmussen_taxonomy_1990} to map the relationships among children’s cognitive skills, CA constraints, tasks, and contexts. Although CWA has a long history in adult‑focused human‑factors research \cite{naikar_cognitive_2017, naikar_consideration_2016, rauffet_designing_2015}, this is—to our knowledge—the first application with young learners. Our analysis identified three primary CA functions-school, discovery, and entertainment—each underpinned by distinct information‑processing flows and interpersonal scaffolds. We also found that children engage most often via phones and tablets (but occasionally on laptops, desktops, or TVs), primarily using Google Assistant and Siri. These findings informed the design of our conversation‑tree recipes, which mirror the scaffolding strategies parents use in homework dialogues. 

\textbf{Study 2} tested whether conversation‑tree “recipes” derived from \textbf{Study 1} improve CA scaffolding at scale to answer \textbf{RQ6:} To what extent do conversation-tree recipes (i.e., structured-prompting) improve knowledge scaffolding across grade levels in CA interactions compared to unstructured, free-form dialogue in large-language models (LLMs)? We prompted GPT 4o‑mini with our recipe based upon structured-prompting (versus a vanilla baseline) on 1,200 simulated child–CA exchanges spanning three modes, four grade bands, three knowledge‑levels, and three temperatures. We measured (a) readability alignment, (b) scaffolded question count, depth, and diversity, and (c) overall coherence against gold‑standard scaffolds. Our quantitative analyses showed that the recipe delivers statistically significant and practically meaningful gains on every scaffolding metric, while temperature had no substantive effect.

Building on both studies, we offer three prioritized design recommendations for child‑oriented CAs: (1) Conversation-trees that mirror parent‑child scaffolding flows; (2) Child‑dedicated profiles to preserve context, track progress, and personalize content; and (3) Content curation controls that empower children and caregivers to select age‑appropriate, culturally relevant resources.

Our work makes the following contributions:

\begin{enumerate}
    \item Cognitive work analysis in Latin American contexts. The first CWA application with Brazilian children, revealing culturally specific constraints, goals, and decision strategies.
    \item Empirical mapping of child–CA information flows. A detailed account of how children solve problems and make decisions across CA functions, used to derive actionable design guidelines.
    \item Integrated framework for CA functions. Evidence that children fluidly shift among educational, discovery, and entertainment uses—and a design to support seamless transitions.
    \item Conversation‑tree recipe for dynamic scaffolding. A script‑light, culturally grounded method for fine‑tuning LLMs to generate grade‑appropriate, pedagogically scaffolded dialogue.    
\end{enumerate}

While our studies focus on Latin American youth, these findings offer broad guidance for designing conversational agents that blend human‑human and human‑computer communication to support effective prompt formulation and scaffolded learning.

The rest of the paper proceeds as follows. Section 2 reviews prior work on child–CA interactions. Section 3 presents our conceptual framework. Section 4 reports Study 1 findings, and Section 5 details Study 2’s quantitative evaluation. In Section 6, we discuss the findings of Study 1 and 2, summarizing design implications. Finally, we conclude the paper with Section 7.

\section{Related Work}
In this section, we first examine how children’s mental models of CAs, and the devices they use, shape interaction patterns. Next, we review educational CA designs, highlighting how prior work has incorporated scaffolding to support learning. We then identify limitations of existing systems and explore emerging approaches for culturally grounded, dynamic knowledge scaffolding. Our aim is to highlight gaps in how children’s information‑processing flows have been translated into CA designs, and to motivate the need for conversation-trees that adapt in real time across academic, discovery, and entertainment tasks.

\subsection{Children's Mental Models of CAs}
Children's mental models of CAs are shaped by their perception that CAs can engage in human-like conversational exchanges and interactions \cite{cheng_scientific_2024}. While children may adopt different conversation modes depending on whether they are interacting with trusted adults or CAs \cite{hiniker_can_2021}, they generally expect their interactions with CAs to mirror those they have with humans \cite{cagiltay_investigating_2020, xu_rosita_2023}.

Children often attribute human traits to CAs, particularly embodied ones. These include traits personality, consciousness, and intentions \cite{garg_conversational_2020, rubegni_dont_2022}. While prior work suggests that younger children (e.g., pre- and early literacy) are more likely to anthropomorphize CAs \cite{druga_hey_2017}, older children (ages 10-12) tend to seek companionship and meaningful social interactions similar to those observed in human-human relationships \cite{cagiltay_investigating_2020}.

Despite these tendencies, children also recognize that CAs are agents designed to simulate human-like interactions. Williams et al. \cite{williams_is_2019} found that children used Theory of Mind (ToM) to assess CAs' capabilities, improving their understanding that CAs were programmed for human-like conversations. Although ToM was not a primary focus in Rubegni et al.'s study \cite{rubegni_dont_2022}, children similarly recognized that embodied CAs had limitations and were not expected to engage in caring relationships.

While these studies provide valuable insights into how children attribute human-like traits to CAs, anthropomorphize them, and gradually learn their limitations, we argue that children’s mental models are also influenced by the goals they aim to achieve through their interactions with CAs, as well as by prior successful experiences. We draw on Craik’s conceptualization of mental models, which describes them as cognitive mechanisms people use to anticipate events and inform problem-solving and decision-making strategies \cite{craik_nature_1943}. We contend that children’s mental models of CAs suggest that child-CA interactions should be designed as a blend of human-computer (i.e., command-response) and human-human conversations. By considering children’s cognitive information processing flows from the perspective of the Cognitive Work Analysis framework \cite{rasmussen_taxonomy_1990}, we can design interactions that support both a blended format of conversational exchanges and the functional needs of children’s mental models.

\subsection{Device Type \& Interaction Patterns}
Prior work demonstrates that the physical and interactive affordances of conversational‑agent (CA) platforms systematically shape how children engage, learn, and collaborate. Several studies find that device modality influences both the functional uses children pursue and the quality of their problem‑solving interactions \cite{lee_dapie_2023, cagiltay_investigating_2020, garg_last_2022}.

\paragraph{Audio‑only vs. multimodal interfaces.}
Smart speakers—lacking visual output—encourage brief, entertainment‑oriented exchanges (e.g., jokes, music) and simple task automation, but they constrain deeper information seeking and independent problem solving due to limited contextual scaffolding \cite{garg_conversational_2020, sciuto_hey_2018, lovato_siri_2015}. In contrast, screen‑based devices (phones, tablets, computers) support richer, goal‑directed dialogue: visual feedback enables multimodal repair strategies, sustained inquiry, and more complex academic activities \cite{xu_rosita_2023, lovato_siri_2015}. However, there is some ground to cover on how these modalities differentially scaffold children’s cognitive processes when tackling educational tasks.

\paragraph{Shared devices and social context.}
CA interactions often occur on shared family devices, which blurs user profiles and can lead to content mismatches or exposure to inappropriate material \cite{le_skillbot_2022, sciuto_hey_2018, garg_conversational_2020}. However, shared usage also provides collaborative problem solving: children use CAs together with parents or siblings during homework, cooking, and play, revealing emergent scaffolding strategies that remain under-characterized in existing work \cite{van_brummelen_what_2023, garg_conversational_2020}. Prior research thus highlights a tension: shared contexts both introduce safety risks and create opportunities for interpersonal support, but lack an integrated account of how collaboration dynamics vary by device type.

\paragraph{Interaction fluency and developmental considerations.}
Young children interacting via voice‑only CAs exhibit frequent mid‑utterance pauses natural in child speech. However, these pauses often trigger premature or incorrect agent responses, disrupting fluency and learning flow \cite{folstad_conversational_2021, cheng_why_2018, xu_exploring_2020}. While these studies identify a need for pause‑sensitive timing models, they do not link such timing challenges to specific device affordances or to children’s evolving information‑processing strategies across tasks.

Although prior work establishes that device affordances influence what children do with CAs (e.g., automation vs. learning) and how they do it (e.g., pausing, co‑use), it remains descriptive and siloed by modality \cite{garg_conversational_2020}. There is no unified framework that (1) maps device‑specific constraints onto children’s cognitive workflows, (2) accounts for the role of more knowledgeable others in scaffolded CA use, nor (3) derives concrete design structures, such as conversation-trees, that can adapt dynamically across devices and collaborative contexts.

The increased interest in designing education-oriented CAs for children suggests that these agents can help children achieve their academic goals. Academic activities usually involve collaborations with teachers, peers and family members, indicating that CAs should support interpersonal assistance. Moreover, children may engage in multiple activities simultaneously \cite{van_brummelen_what_2023}. Thus, our study expands on the types of devices children use to interact with CAs, the levels of collaboration and the interplay across different types of interaction (academic, entertainment/play and task automation).  

\subsection{CAs as Learning Tools}
CAs have shown promise in supporting children's learning by supporting cognitive development, accommodating diverse learning formats, and enabling personalized educational experiences \cite{davidson_designing_202}. Despite recent advancements, challenges persist particularly in replicating the interpersonal dynamics and scaffolding typically found in child-teacher or child-parent interactions \cite{xu_are_2021}. Designing CAs that align with children’s developmental needs and real-world learning experiences remains a critical area of research. This section reviews prior work related to how CAs contribute to children's learning in three key areas: cognitive development, learning formats, and personalization.

\subsubsection{Supporting cognitive development through CAs}
To support cognitive development, we argue that CAs must facilitate not just content delivery, but also encourage children’s critical thinking, reflection, and independent problem-solving. While commercial CAs often fall short in this regard \cite{garg_last_2022}, several research prototypes have incorporated strategies specifically aimed at scaffolding children's cognitive growth. These strategies include timely and personalized feedback, guided question-answering exchanges, contextualized explanations, and prompting critical reasoning during tasks \cite{alaimi_pedagogical_2020, lee_dapie_2023, lovato_hey_2019, xu_exploring_2020}.

Children frequently perceive CAs as social partners, leading to expectations of personalized interactions. Studies have shown that children respond positively to feedback that mirrors interpersonal exchanges, particularly when such feedback is timely and relevant to their learning progress \cite{rubegni_dont_2022, sabnis_empowering_2024}. In most cases, this feedback is implemented through formative assessments that evaluate correctness and provide reinforcement \cite{liu_peergpt_2024, alaimi_pedagogical_2020}. While some systems track children’s performance over time to adapt future prompts \cite{ward_my_2011}, generating open-ended and meaningful feedback that children can fully understand remains a persistent design challenge especially in ensuring that children understand what CAs are communicating in feedback \cite{folstad_conversational_2021}.

Inquiry-based learning is another area where CAs can positively impact cognitive development. Drawing inspiration from parental scaffolding strategies, some CAs guide children to reflect on problems by offering hints rather than direct answers \cite{cheng_why_2018}. This approach promotes active engagement and supports deeper comprehension \cite{lovato_hey_2019, xu_rosita_2023}. Research indicates that when CAs ask children to reflect on their reasoning or explain their answers, children remain motivated and continue engaging with tasks even after meeting minimum completion thresholds \cite{druga_hey_2017, cheng_why_2018, xu_are_2021}.

In addition to supporting reflection, CAs can help contextualize knowledge by linking abstract concepts to children’s lived experiences. Prior work suggests that CAs designed with social cognition features can improve engagement and retention, especially when used to explain complex topics through analogies or relatable scenarios \cite{williams_is_2019, xu_are_2021}. Although primarily explored in higher education contexts, integration with traditional teaching methods such as in-person and distance learning has also been found to enhance meaningful understanding of concepts \cite{winkler_sara_2020}.

Scaffolding problem-solving is another crucial capability for educational CAs. Instead of simply answering questions, well-designed CAs support children by identifying knowledge gaps and guiding them through follow-up prompts or problem-solving pathways \cite{ward_my_2011, garg_conversational_2020, folstad_conversational_2021}. Open-ended questions and personalized inquiries help children explore ideas and articulate thoughts more clearly, promoting sustained interest and deeper learning \cite{alaimi_pedagogical_2020, xu_exploring_2020}. Moreover, educational CAs can be particularly valuable when parents lack the topical expertise to assist with homework. In such cases, CAs can help maintain children’s engagement and emotional regulation during independent learning \cite{cagiltay_my_2023}. Importantly, both parents and teachers have emphasized that educational CAs should encourage comprehension over rote copying, ensuring that children develop meaningful understandings of academic content \cite{han_teachers_2024}.

\subsubsection{Learning Formats in Child-CA Interactions}
CAs have been utilized to support a variety of learning domains, including literacy \cite{lee_dapie_2023, xu_exploring_2020}, writing \cite{cheng_scientific_2024, han_teachers_2024}, and STEM subjects \cite{benotti_engaging_2014, williams_is_2019}. Typically, these agents function as learning companions that guide children through structured tasks by asking content-aligned questions, rather than serving as passive information sources \cite{lee_dapie_2023, monarca_why_2020, xu_exploring_2020}.

The design of learning content plays a major role in sustaining engagement. Davison et al. \cite{davison_working_2020}, for instance, developed an embodied CA that guided children through tangible academic tasks. Although the system maintained engagement during initial use by offering varied and personalized content, engagement dropped when a subsequent task reused the same interface but lacked novelty. This suggests that repeated use of a static interaction design may lead to stagnation, emphasizing the need for adaptive and evolving content structures.

Another critical factor in engagement is the use of multiple media formats. CAs relying solely on text often fail to meet the needs of children with diverse learning preferences \cite{alaimi_pedagogical_2020}. Studies have shown that when CAs incorporate visual, auditory, and interactive content—such as images, videos, songs, or stories—children exhibit higher motivation and retention of knowledge \cite{druga_hey_2017, xu_elinors_2022, xu_rosita_2023}.  Researchers stress that media selection should be purposeful and aligned with learning objectives. For instance, static images may serve to initiate topic exploration, while animations can illustrate dynamic processes or enhance storytelling \cite{ward_my_2011}. The variation in media formats also help children visualize abstract concepts, reinforcing understanding and enabling richer discussion \cite{garg_conversational_2020}.

\subsubsection{Personalizing Learning Experiences with CAs}
Personalization is a defining feature of effective educational CAs. These systems can assess children’s knowledge levels, topical interests, and engagement patterns to adapt both the content and the mode of interaction \cite{agostinelli_designing_2021, bradley_explainable_2022, davison_working_2020}. Adaptive CAs frequently adjust question complexity based on the child's responses and scaffold their learning journey accordingly \cite{ward_my_2011, xu_elinors_2022, winkler_sara_2020, alaimi_pedagogical_2020, han_teachers_2024}.

In addition to academic personalization, children expect that CAs tailor experiences based on their personal interests—such as favourite school subjects, hobbies, or preferred activity types \cite{alaimi_pedagogical_2020, garg_conversational_2020}. However, implementing such personalization, particularly in informal learning settings like home or during play, poses challenges in measuring cognitive load, motivation, and success \cite{ward_my_2011}.

Parental and teacher support for CA integration in academic activities (e.g., homework) depends on whether the agent promotes critical thinking and skill development, rather than simply delivering content \cite{lee_dapie_2023, han_teachers_2024, ho_its_2024}. To support personalization in less structured contexts, a human-in-the-loop model may be a solution. In this approach, parents or educators can  configure the CA’s content delivery, monitor its output, and adjust its recommendations. This model offers a way to balance autonomous learning with informed adult oversight and has potential to bridge gaps between formal and informal learning environments.

\subsection{Conversation Design for Engagement}
Children's preferences for CAs often mirror the conversational styles they experience in classroom settings. They tend to favor interactions that resemble those used by teachers: summarizing, defining, asking questions to assess knowledge, giving topic-related examples, and linking content to real-life scenarios \cite{cagiltay_my_2023}. Research suggests that mimicking teacher-like dialogue structures can enhance children's engagement, motivation, and learning outcomes \cite{garg_conversational_2020, xu_are_2021, xu_rosita_2023}. CAs designed with these principles in mind can recreate interpersonal exchanges that feel meaningful and familiar to children \cite{winkler_sara_2020}.

Prompt formulation is another important aspect of conversation design for CAs. Many children, especially those still developing verbal skills, may need assistance in constructing their queries. CAs that offer guiding questions or rephrase prompts based on the interaction context can support children in articulating their thoughts more clearly \cite{druga_hey_2017, alaimi_pedagogical_2020}. Adapting the tone and structure of prompts—for example, using phrases like “I’m curious about what your answer is”—may sustain engagement while accommodating different cognitive levels \cite {xu_are_2021}.

Open-ended, personalized questions can also encourage children to reflect on their knowledge and learning processes \cite{garg_conversational_2020}. These prompts can allow CAs to collect more contextual information, making it possible to reference past interactions and build upon prior knowledge. However, designing engaging and developmentally appropriate open-ended exchanges remains a challenge, particularly for pre- and early literacy children \cite{folstad_conversational_2021}. Features such as speech transcripts have been proposed to show how the CA interpreted a child’s utterance, which may help support reflection and understanding, though these features work better with older children \cite{xu_are_2021}.

Despite these advances, several limitations persist. Technical constraints such as rigid, scripted flows and insufficient handling of subjective prompts can limit the adaptability of CAs to different learning levels \cite{williams_is_2019, xu_are_2021}, making it challenging to contextualize learning or support progressive knowledge development \cite{ lovato_siri_2015, sciuto_hey_2018}. Moreover, many CAs limit children's agency by failing to encourage independent decision-making or exploration \cite{clark_what_2019, xu_exploring_2020}. These issues highlight the need for a new model of child-CA interactions: one that lies between human-human dialogue and machine-driven command-response structures.

\subsection{Guidance \& Agency Support in CA Interactions}
Many children face difficulties in initiating prompts and recovering from unhelpful responses, largely due to insufficient guidance in formulating their input \cite{han_teachers_2024}. Providing initial support, such as suggested prompts or simplified options, can help children overcome these hurdles and progress in the interaction \cite{han_teachers_2024, lee_dapie_2023, sabnis_empowering_2024, xu_are_2021}. In more responsive systems, the CA can contextualize a child’s input and maintain a fluid back-and-forth conversation that supports learning without providing direct answers \cite{xu_are_2021}.

One promising direction is the adaptation of the \textit{“Questioning the Author”} strategy, which combines animations, images, and interactive questions to stimulate deeper thinking \cite{ward_my_2011}. Though this approach has shown potential, it has so far been tested mainly in controlled settings using \textit{Wizard-of-Oz} prototypes, indicating a need for further development before real-world deployment.

Comprehension challenges are also common among children, especially those in early literacy stages \cite{lovato_hey_2019}. If the language, structure, or complexity of the CA's responses is misaligned with a child’s developmental level, the child may avoid responding altogether or fail to grasp what is being asked \cite{alaimi_pedagogical_2020, cheng_scientific_2024}. Thus, aligning the design of CA prompts and responses with children’s cognitive capabilities is essential for enabling effective learning interactions.

\subsection{Improving CA Speech and Interaction Cues}
Speech-based interaction presents unique challenges for children interacting with CAs. Those with limited verbal skills may speak slowly, hesitate, or pause while thinking, behaviours that current CAs often misinterpret as the end of input, leading to premature or incorrect responses \cite{monarca_why_2020, xu_exploring_2020, le_skillbot_2022}. These limitations are particularly pronounced in question-answering tasks, where children may be checking for comprehension or mentally formulating their response.

To navigate these breakdowns, children often mimic human conversational strategies, such as increasing their volume, repeating prompts, or slowing their speech \cite{cheng_why_2018, garg_conversational_2020, druga_hey_2017}. This behaviour suggests that children internalize responsibility for communication failures, rather than attributing them to the CA \cite{cheng_why_2018}. Moreover, some children interrupt or overlap with the CA’s speech, unaware of appropriate pacing in voice-based interaction \cite{xu_exploring_2020}. Finding the right pace for voice interactions may require some time for children to master as they will likely learn as they gain more experience \cite{sciuto_hey_2018}.

Non-verbal behaviours, such as body movements and facial expressions, are also part of children’s communication repertoire. However, most CAs do not recognize or respond to these cues, leading to confusion and disengagement \cite{xu_exploring_2020}. When the CA fails to acknowledge these non-verbal cues, children do not receive clear guidance on what to do next \cite{liu_peergpt_2024}. This lack of clear communication from CAs is also evident when children struggle to convey their prompts effectively \cite{druga_hey_2017, wang_informing_2022}.

To address these challenges, CAs need to improve their communication cues, ensuring more effective prompt communication, better handling of collaborative interactions, and appropriate interpretation of non-verbal cues.

\subsection{Content Content Curation and Privacy Considerations}
Even when designed for children, CAs can expose users to inappropriate content or raise privacy concerns \cite{le_skillbot_2022}. Designing content moderation mechanisms that are both effective and contextually appropriate remains a major challenge, compounded by the limited availability of child-specific training data, which hampers CAs' ability to accurately interpret children's speech and intent \cite{lovato_hey_2019}.

Transparency in how CAs operate and make content decisions is critical for building trust among both children and caregivers \cite{van_brummelen_what_2023, williams_is_2019}. However, increased trust can sometimes result in overreliance on the CA, with children assuming the agent always provides accurate and safe information \cite{wang_informing_2022}. Enabling children and parents to flag irrelevant or harmful content is one way to increase accountability and safety.

Parental involvement in curating content for CAs may be an alternative to ensure content relevancy and appropriateness. Studies show that parents want to control what content their children can access, including setting filters and scheduling when educational tasks can be performed \cite{garg_conversational_2020, lovato_hey_2019, han_teachers_2024}. Still, many struggle with poorly designed or inaccessible safety features in commercial systems \cite{garg_conversational_2020, sciuto_hey_2018}. Including trusted adults (e.g., parents and teachers) in the content-curation loop, allowing them to define learning goals or limit accessible materials, can address this issue. Yet, decisions about what content children receive often remain in the hands of designers, not families or educators \cite{alaimi_pedagogical_2020, cagiltay_my_2023, garg_last_2022, lee_dapie_2023}.

\subsection{Profile Personalization for Child-Centric Interactions}
Personalization is another critical area in child-CA design. Differentiating between child and adult users may enable the system to deliver developmentally appropriate content and interaction styles \cite{lovato_siri_2015, ho_its_2024}. Features such as voice-based self-identification or dedicated user profiles can help tailor the experience accordingly \cite{lovato_hey_2019, ho_its_2024}.

Children use CAs in diverse contexts, ranging from academic tasks to entertainment and household routines, and expect the system to adapt accordingly. They want autonomy over what tasks the CA performs and how interactions are personalized based on their interests \cite{garg_conversational_2020, rubegni_dont_2022}. This expectation emphasizes the need for flexible, task-aware CAs that recognize the child’s role and context of use.

Parents also expressed a desire to periodically revise learning objectives and adjust settings to reflect their child’s evolving academic progress \cite{ho_its_2024}. However, most current commercial CA interfaces lack intuitive controls for managing these preferences, limiting parent involvement in content curation For example, parents often report difficulties in setting up safety features in commercial CAs or even being unaware of their existence, indicating that these features are not user-friendly \cite{le_skillbot_2022, sciuto_hey_2018}. Furthermore, personalization features raise concerns about data collection and use. Parents seek transparency about what data is collected, how it is used, and for what purpose \cite{garg_conversational_2020, ho_its_2024}. Designing CAs to clearly communicate their data practices is a crucial step toward building trust and safeguarding children’s digital experiences. Another aspect remains in adjusting CAs conversations to support local cultures and pedagogies \cite{locatelli_examining_2025, ogan_collaboration_2012}.

\subsection{Culturally-Driven CA Interactions}
Most commercial CAs are designed and culturally tuned for Western, English-speaking markets \cite{danielescu_bot_2018, lugrin_adapted_2018}. Simple translation of their dialogue scripts into other languages (e.g., Brazilian Portuguese) often produces stilted exchanges that ignore local idioms, politeness norms, and social power dynamics \cite{chaves_chatbots_2022}. In Latin America, where device access, network reliability, and educational infrastructures vary dramatically, these “one-size-fits-all” CAs can feel inauthentic or even alienating \cite{salgado_human_2015, portela_ai_2024}. Moreover, fixed agent personalities (e.g., voice, humour style, turn-taking patterns)rarely adapt to the collectivist orientation and high-context communication styles common in many Latin American households and academic contexts \cite{ogan_collaboration_2012}. As a result, children disengage when conversations fail to reflect their cultural expectations.

HCI researchers have explored rule-based and co-design methods to inject local flavour into CA dialogues. During a participatory design workshop, Brazilian adolescents designed politeness formulas and turn-taking cues that mirrored interactions characteristic to interpersonal exchanges in Brazilian culture \cite{monteiro_investigating_2024}. In another study, emotion-recognition models fine-tuned on child-specific facial datasets showed not only how culturally grounded nonverbal cues can be integrated, but also that it improved CA's ability to interpret nuanced emotional cues \cite{zimmer_hybrid_2024}. Yet these approaches can remain siloed: rule-based scripts or narrow domain models that cannot generalize beyond their original context.

Large language models (LLMs) offer a scalable path beyond static scripts. Fine-tuning GPT-4 or open-source LLMs (e.g. Sabiá, MariTalk) on Brazilian-specific corpora (e.g., national exam essays, children’s stories, classroom transcripts) can yield  more culturally relevant outputs in humanities tasks (e.g., language and social sciences) that hinge on cultural nuance, than natural sciences whose topics can be generalized across cultural contexts \cite{locatelli_examining_2025}. By training on region‑specific datasets, LLM‑powered CAs fined-tuned with culturally relevant datasets can deliver culturally relevant explanations, adapt to local pedagogies, and maintain topical and social appropriateness.

Moreover, educational technology infused with LLMs can generate a positive impact on student’s learning experiences, particularly in self-learning, specially for those living in impoverished areas or areas with limited educational infrastructure when teachers cannot be present in the classroom. While LLM-powered CAs offer a promising pathway, socioeconomic inequities in Latin American contexts limit stable internet connectivity and device availability \cite{alvarado_fostering_2020, ogan_collaboration_2012, salgado_human_2015}. Offline‑capable, AI‑infused tools, such as the offline LLM prototype designed by Portela et al. for personalized writing feedback on low‑cost hardware—demonstrate how CAs can reach underserved students without disrupting existing classroom practices \cite{portela_ai_2024}.  However, accessing such contexts present challenges due to participant-researcher biases, language barriers and, even, the availability of technology infrastructure to deploy such technologies \cite{portela_bringing_2024, salgado_human_2015}.

Hybrid human–machine dialogue design must account for culturally driven interaction norms. For instance, comparative studies between U.S. and Latin American students (Brazil, Costa Rica, Mexico) reveal differences in power distance, tolerance for ambiguity, and individualism–collectivism orientations \cite{ogan_collaboration_2012}. Whereas North American learners tackle cognitive tutors independently, Latin American students often collaborate with peers and await explicit teacher direction before engaging with intelligent agents, a pattern rooted in collectivist family and classroom dynamics. These findings suggest that LLM-powered CAs should embed clear turn-taking cues and scaffolded instructions that signal when it is the child’s opportunity to respond or act. Moreover, conversation designs for CAs must support multi-party scaffolding, inviting peer or adult contributions, and adapt the depth and explicitness of prompts to local pedagogical practices. By modeling dialogues as a distinct human-machine category with persistent context, dynamic turn management, and culturally tuned scaffolds, designers can create CA interactions that feel as natural and supportive as in-person learning exchanges.

\section{Conceptual Framework}
We adopt a quasi-ecological perspective to analyze child-CA interactions. This perspective views human interactions with technology as part of an interconnected system that supports recurring interaction and behavioural patterns \cite{fidel_human_2012, rasmussen_taxonomy_1990}. Furthermore, a quasi-ecological perspective considers people's cognitive capabilities and personal experiences when analyzing child-CA interactions. In the following sub-sections, we will discuss relevant theoretical foundations for designing child-oriented CAs, providing a basis for discussing the application of a Cognitive Work Analysis framework in our study.

\subsection{Theoretical Foundations for Designing Child-Oriented CAs}
The design of child-oriented conversational agents (CAs) can be grounded in developmental and educational psychology, which provides critical insight into how children learn, think, and interact. These theoretical foundations guide the creation of CAs that do more than deliver information—they act as cognitive partners, scaffolding children’s learning and supporting autonomy, critical thinking, and reflection.

A central concept in this domain is that of the more knowledgeable other (MKO), drawn from Vygotsky’s sociocultural theory of learning \cite{vygotsky_mind_1978}. Children construct knowledge through guided interaction with MKOs (e.g., parents, teachers, friends and family members), who provide support by calibrating questions, hints, and feedback to match the child’s developmental stage. When applied to CAs, this framework positions the agent as a learning facilitator capable of adapting to the child’s abilities and goals.

From a Cognitive Work Analysis (CWA) perspective, designing CAs that serve as MKOs requires an understanding of how children engage in concurrent tasks such as learning, entertainment, and automation. CAs must balance the precision of human-computer interaction models with the fluidity and contextual depth of human-human conversation. This hybrid interaction model is essential to support knowledge scaffolding, independent reasoning, and developmentally appropriate engagement.

Several theories provide actionable guidance for achieving this. Agency, for example, refers to the learner’s capacity to make decisions and direct their own learning processes \cite{bandura_toward_2006}. Children show greater motivation and engagement when they have control over learning choices. CAs can support agency by allowing children to set goals, select activities, and decide how to approach tasks like homework, which fosters ownership and supports self-directed learning \cite{lee_dapie_2023}. While younger children may sometimes gravitate toward simpler tasks, well-designed CAs can help educators and parents identify and address learning barriers while preserving the child's sense of control.

Theory of Mind (ToM)—the ability to recognize that others have distinct thoughts and perspectives—is also relevant to child-CA interaction \cite{premack_does_1978}. Children often attribute human-like qualities to CAs and expect them to interpret social and emotional cues appropriately \cite{marchetti_theory_2018}. However, since ToM is not inherent to CAs, designers must compensate by prompting children to explicitly express their thoughts and emotions. This can be achieved through dialog strategies that simulate perspective-taking and by enabling CAs to respond based on accumulated contextual knowledge. Such interactions align with children's expectations and support deeper engagement through personalized, inquiry-based conversation.

Building on this, Vygotsky’s concept of the Zone of Proximal Development (ZPD) highlights the importance of adaptive support, helping children achieve tasks just beyond their independent abilities. CAs can operate within a child’s ZPD by dynamically adjusting their level of scaffolding based on the child’s inputs, performance, and confidence. This ensures that challenges remain within reach while still promoting growth, a strategy shown to facilitate learning and cognitive development \cite{alaimi_pedagogical_2020}.

Lastly, the Questioning the Author (QtA) approach offers a practical framework for encouraging children to interrogate and make sense of information \cite{beck1996questioning}. QtA emphasizes the importance of reflection and knowledge integration by prompting learners to question sources, connect new content to prior knowledge, and explore meaning. Applied to CAs, QtA-inspired conversation flows have been shown to foster critical thinking and deeper engagement \cite{ward_my_2011}. When children are invited to evaluate the information provided by a CA, rather than passively receive it, they are more likely to develop analytical skills and maintain curiosity throughout the learning process.

Together, these frameworks—Agency, ToM, ZPD, and QtA—establish a foundation for designing child-oriented CAs that can serve as adaptive, engaging, and socially responsive learning companions. Moreover, these frameworks provide a basis for investigating children's information processing flow in interactions with CAs.

\subsection{Cognitive Work Analysis Framework}
Building on theoretical foundations that position CAs as cognitive partners, we adopt a Cognitive Work Analysis (CWA) perspective to explore how children interact with CAs across diverse contexts. CWA offers a systems-oriented approach that examines how tasks, environments, and user characteristics shape interaction dynamics through a network of constraints.

By analyzing child-CA interactions within this interconnected system, we can better understand both the limitations children face and the strategies they develop to navigate them. While constraints often reveal usability challenges and design gaps, children's adaptive behaviours, such as how they problem-solve, seek help, or persist through breakdowns, offer critical insights for designing interfaces that better align with their cognitive and developmental capabilities.

CWA focuses on a range of constraints, including contextual factors such as setting, activity rules, available tools and resources, and individual skills and mental models \cite{vicente_cognitive_1999}. These elements do not act in isolation; rather, they influence one another and collectively shape how children interact with technology to complete their goals. As users navigate these constraints, they continuously adjust their actions and decisions, revealing patterns of behaviour that emerge from the broader sociotechnical system \cite{rasmussen_taxonomy_1990}.

To guide our analysis, we draw on Rasmussen’s CWA framework \cite{rasmussen_taxonomy_1990}. This tool supports the examination of child-CA interaction by mapping out relationships between high-level goals, the means available to achieve them, and the constraints that shape these possibilities. The CWA consists of five:

\begin{itemize}
    \item \textbf{Work Domain Analysis} maps the means-ends structure of the interconnected system.
    \item \textbf{Organization Analysis} identifies collaborations, role allocation, and social organization.
    \item \textbf{Control Task} explores individual and collaborative activities and situations in work domain terminologies, decision-making ladder, and strategies and heuristics.
    \item \textbf{Strategies Analysis} identifies different ways operators can accomplish tasks under varying conditions, emphasizing flexibility and adaptability.
    \item \textbf{Actor Analysis} focuses on individual skills, domain knowledge, and cognitive capabilities.
\end{itemize}

In the CWA, we distinguish between settings, situations, and scenarios. Settings refer to the physical environment where activities occur, while situations involve the typical routines intrinsic to a role (e.g., student). Scenarios, on the other hand, encompass the range of possibilities that can arise in certain situations given the circumstances in which people find themselves. For each situation, people delineate information-processing flows to complete activities, which are sets of tasks. These information-processing flows represent various courses of action that can be taken in a given situation, based on the interplay of prior domain knowledge, individual skills, and predefined rules (e.g., policy, corporate, role-specific, and technical). 

The means-ends analysis (MEA) in the CWA connects higher-level goals and purposes (ends) with the means available to achieve them, organizing findings into five hierarchical abstraction levels:
\begin{itemize}
    \item \textbf{Goals/Constraints} concern the overarching objectives, opportunities and limitations guiding an activity.
    \item \textbf{Priorities} determine what needs to be addressed first based on urgency and significance.
    \item \textbf{Functions} refer to the purposes users attribute to devices and applications (e.g., CAs).
    \item \textbf{Processes} are the detailed procedures or workflows followed to carry out the activities and tasks part of these activities.
    \item \textbf{Resources} are the tools, materials, or supports used to perform the processes, encompassing items or information needed to accomplish the tasks.
\end{itemize}

CWA’s holistic approach provides a detailed understanding of the interconnected constraints shaping child-CA interactions. By examining the environment, activities, and interactions, we can pinpoint specific factors that enable or hinder effective CA use. MEA's hierarchical structure clarifies how children's goals and purposes with CAs are supported or limited by available resources and constraints. This comprehensive analysis is crucial for understanding the complexity of child-CA interactions and for identifying areas where CAs can be improved to better support children's needs.

\section{Study 1 — Child-CA interactions}
\subsection{Methods}
The research discussed in this paper was part of a broader investigation into the use of digital applications in children’s homework information-search routines, with a focus on children's interactions with conversational agents (CAs) in everyday and academic activities. We obtained approval from the university's institutional review board to ensure ethical compliance.

We conducted an exploratory online study with 23 participants, including children, parents, and teachers, between May and July 2021. While COVID-19 restrictions remained in effect in some regions, their impact on school attendance varied among participants. Of the nine children in our study, two (C04 and C09) were enrolled in a hybrid learning format, attending in-person classes every other week. The remaining seven children attended regular in-person classes full-time. All participating teachers were teaching regular in-person classes full-time.

To accommodate participants' preferences and reduce technical barriers, we asked them to choose the video conferencing tool they felt most comfortable using. Consequently, we conducted the study using Zoom, Google Meet, WhatsApp and Skype.

Our quasi-ecological study employed a multi-stage, multi-modal approach, encompassing several study sessions and multiple methods. In the following subsections, we describe the study design for each participant group. These sections are organized by participant type (e.g., children/parents or teachers), and within each, we outline the methods and recruitment process used in this study.

\subsubsection{Participants}
We recruited nine Brazilian children (ages 9-11) and their parents. Recruitment occurred in two phases (Table \ref{tab:child-parent-demographics}). In the first phase, a mediator in Brazil distributed the study invitation letter via a WhatsApp group consisting of 32 caregivers, primarily parents of Year 4 students at an elementary school in the Rio de Janeiro metropolitan area. This school was selected for its diverse student backgrounds. The mediator, a parent of a Year 4 student, collaborated with the group administrator (a Year 4 teacher) to share the invitation. The letter outlined the study’s purpose, participation requirements, and instructions for expressing interest. Four mothers initially responded and took part in an introductory session. In the second phase, these mothers invited other interested caregivers, leading to the participation of five additional mothers. Although gender was not an inclusion or exclusion criterion, only mothers responded to the study invitation. Each participant received a \$10 gift card per completed session, up to \$70 for seven sessions.

\begin{table*}
  \caption{An overview of demographics of the nine child-parent pairs (18 participants) in this study. Each child (C) is paired with their corresponding parent (P), where the number following the child's ID indicates their age (in years), and the description after the parent's ID represents their education level. For example, C01 (age X) is paired with P01 (education level).}
  \label{tab:child-parent-demographics}
  \begin{tabular}{ccccl}
    \toprule
    Participant type & Participant ID &Gender &School & Video conferencing tool\\
    \midrule
    Child & C01 (9) & Female & \multirow{2}{*}{S01} &\multirow{2}{*}{Zoom}\\
    Parent & P01 (Post-secondary) & Female & & \\
    \midrule
    Child & C02 (9) & Male & \multirow{2}{*}{S01} &\multirow{2}{*}{Zoom}\\
    Parent & P02 (Secondary) & Female & & \\
    \midrule
    Child & C03 (10) & Female & \multirow{2}{*}{S02} &\multirow{2}{*}{WhatsApp}\\
    Parent & P03 (Elementary) & Female & & \\
    \midrule
    Child & C04 (9) & Female & \multirow{2}{*}{S01} &\multirow{2}{*}{Zoom}\\
    Parent & P04 (PhD) & Female & & \\
    \midrule
    Child & C05 (9) & Male & \multirow{2}{*}{S03} &\multirow{2}{*}{Zoom}\\
    Parent & P05 (Post-secondary) & Female & & \\
    \midrule
    Child & C06 (10) & Male & \multirow{2}{*}{S04} &\multirow{2}{*}{Zoom}\\
    Parent & P06 (Post-secondary) & Female & &\\
    \midrule
    Child & C07 (9) & Male & \multirow{2}{*}{S05} &\multirow{2}{*}{Zoom}\\
    Parent & P07 (Post-secondary) & Female & &\\
    \midrule
    Child & C08 (9) & Male & \multirow{2}{*}{S01} &\multirow{2}{*}{Zoom}\\
    Parent & P08 (Post-secondary) & Female\\
    \midrule
    Child & C09 (9) & Male & \multirow{2}{*}{S06} &\multirow{2}{*}{Skype}\\
    Parent & P09 (Elementary) & Female & &\\
    \bottomrule
  \end{tabular}
\end{table*}

We also recruited five teachers (Years 4-5). Teacher recruitment was primarily through snowball sampling (Table \ref{tab:teacherdemographics}). The Year 4 teacher (T01), who managed the WhatsApp group, shared the study invitation with colleagues at the same school and other schools. An information technology (IT) teacher (T02) teaching Years 1-9 at the same school as T01 responded, as did three other teachers from different schools—two Year 4 teachers (T03 and T04) and one Year 5 teacher (T05). T01 and T02 taught classes to students C01, C02, C04, and C08. Although we did not consider gender as an inclusion or exclusion criterium, most teachers interested in participating in our study were female. Each participating teacher received a \$50 gift card as compensation for their involvement in the study.

\begin{table*}
  \caption{An overview of the demographics of the five teachers in this study. Each teacher (T) is identified by an ID, with the number following it indicating the grade level(s) they taught. Teachers T01 and T02 instructed children C01, C02, C04, and C08, while the remaining teachers did not teach any of the children in this study. Most teachers taught all subjects, except for T02, who specialized in IT.}
  \label{tab:teacherdemographics}
  \begin{tabular}{ccccl}
    \toprule
    Teacher ID &Gender & School &Subject & Video conferencing tool\\
    \midrule
    T01 (Grade 4) & Female & S01 & All & Zoom\\
    \midrule
    T02 (Grades 1-10) & Male & S01 & IT & Zoom\\
    \midrule
    T03 (Grade 4) & Female & S07 & All & Zoom\\
    \midrule
    T04 (Grade 5) & Female & S08 & All & Zoom\\
    \midrule
    T04 (Grade 5) & Female & S09 & All & Google Meet\\
    \bottomrule
  \end{tabular}
\end{table*}

\subsubsection{Child-Parent Procedure}
We conducted a seven-week online multi-modal, multi-stage study. Each child-parent pair participated in an introductory session, an interview, an online observation, three critical incident sessions, and a reflective interview. All sessions were video recorded, except the introductory session and the reflective interview. Oral consent was obtained at the beginning of every session.
We applied the methodology described in \cite{figueiredo-2024-explora} to conduct the studies with children and parents, deliberately structuring the method sequence to optimize the study design.

\textbf{Introductory Session (20-30 minutes).} The introductory session served as the first point of contact between the researcher and participants. We used this unrecorded meeting to build rapport, explain the study, demonstrate CA use (e.g. “Check Rio’s weather”), address questions, obtain parental consent and child assent, and schedule subsequent sessions.

\textbf{Interview (30-45 minutes).} The interview aimed to identify the participants’ context, understand how they make sense of their interactions with CAs, and determine their preferences and constraints regarding CAs. We interviewed the child first to prevent the parent’s perspective from influencing their responses. The \textit{"can you teach me [this]?"} prompt was used to help children articulate their thoughts. The interviews were video recorded, and notes were taken.

\textbf{Online Observation (40-60 minutes).} Prior to each session, we reviewed interview notes to pinpoint likely CA‐relevant homework tasks and scheduled a convenient time. Using the family’s preferred video‐conferencing tool, children and parents set up their camera in the usual homework space (e.g. bedroom or kitchen) and adjusted it as needed if they moved outside the frame. We prompted children with \textit{“Can you teach me how you would complete this activity?”} \cite{figueiredo-2024-explora} to elicit natural interactions, then followed up on observed strategies, decision points, and context. This remote approach, necessary under COVID-19, allowed us to capture authentic behaviour, environmental cues, and the reasoning behind children’s CA use without disrupting their routine.

\textbf{Critical-Incident Sessions (40-60 minutes).} Over three sessions, children and parents recalled and reenacted prior CA interactions—sometimes by completing real homework—using the same \textit{“Can you teach me…?” prompt}. A critical incident is a noteworthy event that reveals key successes or failures in a system or interaction \cite{weatherbee2010critical}. Journals children kept of their CA use helped ground these reenactments. The aim was to validate and deepen our understanding of recurring interaction patterns.

\textbf{Reflective Interview (30-45 minutes).} In this session, we revisited video excerpts and presented our preliminary information-processing flows to children and parents, inviting them—via a teacher–student metaphor—to critique and refine our assumptions. To minimize discomfort and bias, we took detailed notes instead of recording, using their feedback to correct and enrich our interpretations.

\subsubsection{Teacher Procedure}
We conducted 30-60 minute interviews with teachers (Years 4-5). Approximately one week later, each teacher participated in a photo elicitation session.

\textbf{Interview (30-60 minutes).} After a brief warm-up, we asked teachers about their classroom context, their experiences with children using CAs for learning, and how they characterize those interactions. We prompted explanations with \textit{“Can you teach me [this]?”} to elicit detailed examples.

\textbf{Photo Elicitation Session (30-45 minutes).} Teachers photographed classroom scenes involving CAs (3–14 images each, M = 8.2) and shared them via WhatsApp. During audio-recorded discussions, they described each image’s context and relevance to CA use.

\subsection{Data Analysis}
We collected approximately 30 hours of recorded data (children and parents: \,\(\approx 23\)\ h; teachers: \,\(\approx 7\)\ h), including video, audio, notes, memos, and photographs (teachers). The analysis steps included data cleaning, transcription, translation, and thematic analysis.

\textbf{Transcription and Translation.} Transcriptions and translations were completed within one week of data collection to refine notes and connect them with specific data examples. Timestamps were added to video recordings for relevant quotes and interactions.

\textbf{Thematic Analysis.} We employed deductive thematic analysis \cite{braun_using_2006}, using the Cognitive Work Analysis (CWA) framework \cite{vicente_cognitive_1999} as our \textit{a priori} themes. Our goal was was to map children’s information‑processing flows, goals, constraints, priorities, and resources, and then translate those insights into design artifacts (e.g., conversation‑flow templates) for child‑oriented CAs. By grounding our deductive thematic analysis in the CWA framework, we ensured analytical validity and reliability and therefore did not perform additional inter‑rater reliability checks.

\textbf{Reflective Interview.} Preliminary analysis focused on notes and memos from each session to seek participant validation during the reflective interview. We began with data from the reflective interview, comparing our preliminary assumptions with participants' assessments. We analyzed the corrected list of items and notes from these sessions.

\textbf{Interview.} Preliminary analysis of interview notes highlighted participants' terminologies, immediate context definitions, and situations involving CA interactions. This informed subsequent sessions by identifying relevant contextual aspects. Interview data were transcribed verbatim and analyzed using deductive thematic analysis.

\textbf{Observation and Critical-Incident Sessions.} Data included video and screen recordings of interactions. Verbal exchanges were transcribed verbatim, and screen movements were documented in square brackets within a single transcription file. We then applied deductive thematic analysis to these transcriptions.

\textbf{Photo Elicitation Session.} Analysis relied on teachers' descriptions of the pictures. To minimize researcher bias and capture naturalistic perspectives, we focused on teachers' interpretations of the scenes depicted in the photographs. This approach, structured around CWA/MEA, allowed us to compare stakeholders' perceptions and experiences of child-CA interactions, emphasizing their roles in these interactions.

\subsection{Findings}
\begin{figure}
    \centering
    \includegraphics[width=\linewidth]{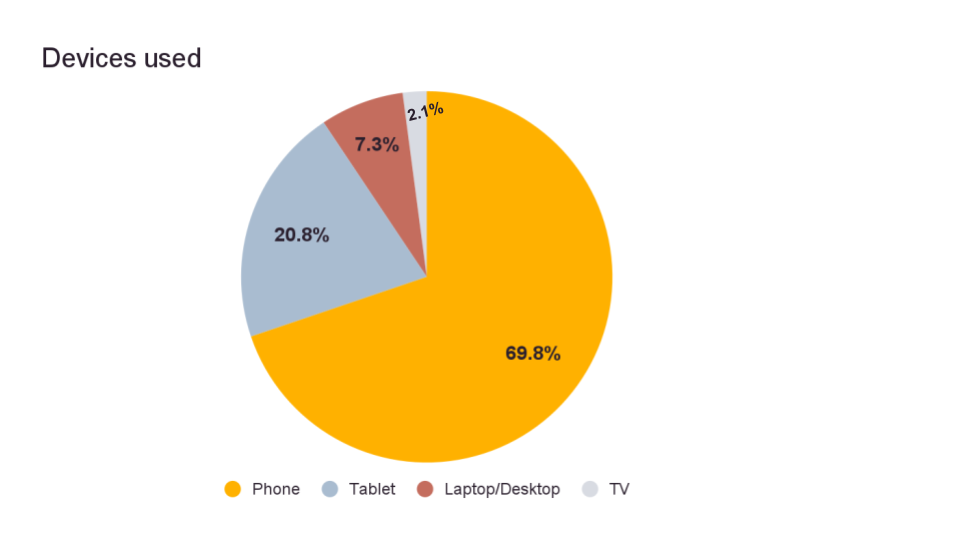}
    \caption{The distribution of devices children used to interact with CAs in Study 1.}
    \Description{This pie chart illustrates the distribution of devices children used to interact with conversational agents (CAs) in the study. The majority of interactions occurred through phones (70\%), followed by tablets (21\%), computers including desktops and laptops (7\%), and TVs (2\%).}
    \label{fig:devicesused}
\end{figure}

We examined 96 natural child–CA interactions in Brazilian homes, capturing every step from \textit{“Hey Google/Siri…”} to finding (or failing to find) needed information. Most interactions happened on phones (70 \%) and tablets (21 \%), with few on computers (7 \%) or TVs (2 \%) (Figure \ref{fig:devicesused}). Google Assistant dominated (6 children), Siri was used by one (C03), and two children (C04 and C08) alternated between both.

Although child-CA interactions started with an awakening voice command, children did not use this command when resuming and restoring interactions. Resumed interactions referred to when children interrupted their interaction with CAs to perform other tasks (e.g., consult a more knowledgeable other (MKO), copy homework solutions to textbooks, check other apps), and then returned to the CA to complete the remainder of the activity, using dialogues like \textit{“Find me stories about deep-sea creatures.”} (C04). In these situations, children perceived the CA as constantly available in the background, waiting for the next command. Restored interactions referred to attempts to restore prior interaction sessions. These prior interaction sessions had happened at least two hours before, with some of these having occurred days before. Parents recalled \textit{"if homework is too difficult, he'll need a couple of days to finish it. So, we try to find everything again"} (P07). 

\subsubsection{Information-processing flow in child-CA interactions}
Children treated CAs as if a real person were on the other end, listening, researching, and responding. C09 said that \textit{"When I ask things to Google Assistant, a person listens, looks up what I'm asking and reads me the answer"}, while another (C06) felt like \textit{ "I'm chatting with my friend over the phone."} They expected CAs to remember past interactions, understand their preferences, guide their problem-solving, and even ask clarifying questions: \textit{“I think the Google lady could help me solve my math homework. Sometimes, my mom doesn’t know how to solve it.” (C02)}. C07 went so far as to hope that \textit{“Google Assistant should replace the person there for someone like Ms. K (C07's teacher) on the other side, so I could ask questions and she’d help me.”}.

Yet this mental model sometimes led to breakdowns. Children expected the CA to tolerate natural pauses while they read homework prompts aloud, but the CA often cut in prematurely:

\begin{quote}
    \textit{“Hey Google, what is a po… wait a second (C05 opens the textbook and reads the homework description, taking about 40 seconds)} (C05)\par
    \textit{Pó (dust) is a fine, dry powder consisting of tiny particles of earth or waste matter” (Google Assistant)} \par
    (C05 overlaps with Google Assistant)\par
   \textit{ “No, stop. I didn’t say what I want”.} (C05)\par
\end{quote}

Children believed that CAs could perceive that they had changed their minds or their contextual cues, similarly to what it happens in interpersonal conversations.

Beyond homework, children saw CAs as companions, asking about their “favourite food” or wishing \textit{“Siri would guess what I liked and showed me exactly what I wanted”} (C04)—treating them like new friends rather than automated search tools or digital assistants, like in the example below: 

\begin{quote}
    \textit{“What’s your pet?”} (C09)\par
    \textit{“My pet is a banana (Google Assistant laughs)}\par
    \textit{"What?! This is strange. No one has a banana for a pet.”} (C09)\par
    \textit{“You’d never understand what it is like to live on the cloud.”} (Google Assistant)\par
    \textit{“She lives on the cloud? She’s always tricking me.”} (C09)\par
\end{quote}

Children organized their information processing flows based on communication cues from CAs, including system feedback (e.g., auditory, visual, haptic, and task completion) and dialogue flows. Viewing these interactions as akin to human conversations, children and their parents expected the CAs to reason, contextualize, and interpret both visual cues (e.g., facial expressions) and auditory cues (e.g., silence). To capture how these expectations shape their step-by-step processing, we modelled children’s flows at three levels (Figure \ref{fig:information-processing}):

\begin{figure}
    \centering
    \includegraphics[width=0.8\linewidth]{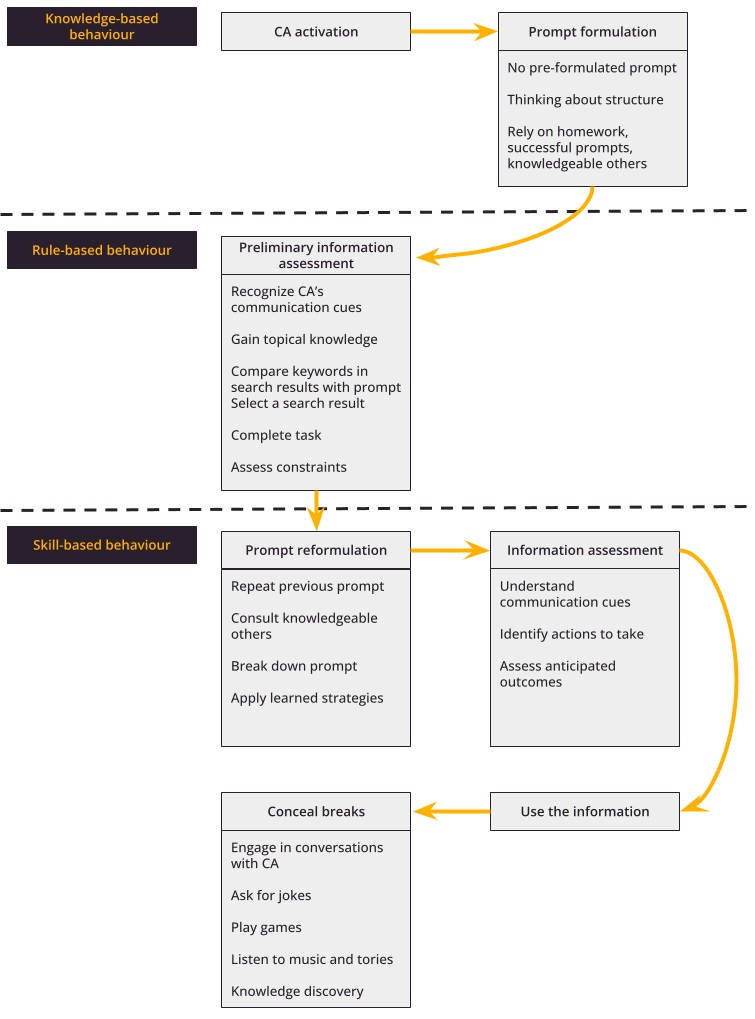}
    \caption{Children's information processing flow emerging from interactions with CAs. Knowledge-based, rule-based and skill-based behaviours represent the three levels of control of child actions. The arrows indicate the information flow, but in real scenarios, these flows can occur simultaneously.}
    \Description{description}
    \label{fig:information-processing}
\end{figure}

Children’s prior experiences with technology and similar applications shaped their knowledge-based behaviour. Familiarity with search engines, family conversations, and digital toys led children to use long, complex prompts, expecting the CA to provide contextually relevant and engaging responses. For instance, one child remarked, \textit{"It feels like Google, but I don’t need to type anything."} (C07). Despite their familiarity with these technologies, all children and four parents (P02, P03, P06 and P09) often struggled to critically evaluate the CA's responses due to a limited understanding of how CAs operate, resulting in an over-reliance on its outputs. \textit{“I know it’s wrong, but I don’t check the information searched when we’re just fact-checking things.”} (P02)

Children’s skills with various technology types, including search engines and voice interfaces, affected their understanding and utilization of communication cues from CAs. For instance, C04 said, \textit{"I use AprendiZap, and it has 'sections' in it, so I don't have to read everything. Why doesn't Google Assistant have that?"} while C03 noted, \textit{"It's much easier to listen to what Google Assistant is saying. I don't have time to read everything."} Despite valuing succinct information, children sometimes misunderstood voice summaries as comprehensive, which did not foster critical thinking. They assumed that the information provided was sufficient for their needs and would meet their teachers' expectations. When CAs did not guide them effectively, children encountered trial-and-error scenarios, repeatedly using or breaking down prompts to find successful outcomes, as seen when C02 mentioned, \textit{"Google Assistant took me to the wrong place."}

\subsubsection{More knowledgeable other's roles in children’s CA-supported problem solving}
Children’s interactions with CAs rarely occurred in isolation: they routinely collaborated with parents, teachers, or peers—“more knowledgeable others” (MKOs)—to frame problems, craft effective prompts, and interpret CA responses. This collaborative scaffolding shaped when children asked for help, which strategies they tried first, and how they adapted when CAs faltered.

Four parents (P01, P06, P08, P09) worked side-by-side with their children throughout the homework process. They began by prompting reflection as illustrated in the example below:
\begin{quote}
   \textit{ “Go and read the homework description. Then, tell me what you got from that.”} (P09)\par
  \textit{  “I think it’s about baroque music. I need to find examples of baroque music.”} (C09)\par
   \textit{ “How would you find those examples?”} (P09)\par
   \textit{ “I think it’d just ask CA ‘give me examples of baroque music.”} (C09)\par
    \textit{“Go ahead and ask it.”} (P09)\par
\end{quote}

Children were asked to understand the homework, devise solutions, and articulate their approach to parents. If difficulties arose, parents intervened with scaffolding questions, offering hints aligned with the homework's demands, allowing children to opt for strategies they deemed effective
\begin{quote}
    \textit{“See. It’s a lot, isn’t it? What’s described in the homework description?”} (P09)\par
   \textit{ “Write three examples of baroque music.”} (C09)\par
    \textit{“What do you think that is? Name of composers…”} (P09)\par
    \textit{“Oh, name of baroque songs?”} (C09)\par
    \textit{“Maybe, why don’t you give it a try?”} (P09)\par
    \textit{“Find the name of three baroque songs.”} (C09)\par
   \textit{ “Isn’t that a little bit better?”} (P09)\par
\end{quote}

Children assimilated these demonstrations, later applying learned strategies autonomously, \textit{“I’m asking this to Google Assistant first to understand what this means, like you (P01) said I should do last time.”} (C01).

Two parents (P04 and P07) encouraged their children to independently solve homework. In this situation, the parents established homework rules, \textit{“Read the homework description and use the CA to search for that information. After you find the answer, I’ll check it.”} (P04). This approach encouraged children to work on their own first, understand the homework problem, find a solution, and explain their problem-solving process back to the parents:
\begin{quote}
   \textit{“I selected these three Brazilian Indigenous artifacts because that was in the homework description.”} (C04)\par
   \textit{“I see, that’s correct. But don’t you think that there’s something missing? You need to describe these artifacts using your own words.”} (P04).\par 
\end{quote}

In situations like the above, parents provided direct directions for homework solutions as their children had dedicated some time prior to solving those homework activities. P07 explained, \textit{“I like to let him understand the homework requirements and try to come up with something on his own. Most of times, I’ll just point to the remaining portion of the solution if what he found doesn’t have the entire solution. Sometimes, the homework description is too vague or has words that aren’t easy to understand.”}

A third group of parents (P02, P03 and P05) preemptively resolved homework complexities for their children. They recognized barriers like understanding abstract concepts or selecting appropriate information sources, devising digestible homework plans tailored to their child's comprehension level and academic struggles. This proactive approach included prior assessment based on homework structure, child's topical knowledge, and academic performance, \textit{“If I don't select the sentences so that he can ask the CA or do the search on his behalf, he won’t make any progress. He’ll get bored and disengaged. I know teachers may find it wrong, but I think it helps him get things started.”} (P02). T03 mentioned that \textit{“parents that are more involved actually help us in assessing what kids require more attention. So, I can tweak the assignment difficulty one bit up or down or make the way I teach the topic easier to understand.”} Moreover, these three parents searched the CA for their children, explaining what words to use:
\begin{quote}
    \textit{“Look, this is about the Pero Vaz de Caminha’s letter, but we need to have the letter first, then look it up in the letter how he described the landscape. ‘Hey Google, where can I find the Pero Vaz de Caminha’s letter?’"} (P02)\par
    (Google Assistant returns the links)\par
   \textit{ "See, here’s the letter (P02 points at the phone screen, showing C02 the letter). We wouldn’t find anything if we had asked ‘landscape description in Pero Vaz de Caminha’s letter’.”} (P02) \par
\end{quote}

Teachers complemented parental support by providing structured guidelines for evaluating CA outputs and translating assignments into CA-friendly keywords. For open-ended projects, they encouraged student agency (\textit{"must create (their) own designs rather than replicating (the) examples showed in class", T01)}, and even supplied sample search phrases like T01: \textit{“I have examples to help my students establish a reference, but I emphasized they needed to find other toys. I even included keywords for them to search on CAs.”} For more procedural tasks, they curated reliable resources but noted the challenge of converting resource titles into effective voice commands:
\begin{quote}
    \textit{“Hey Google, search for Escola Games.”} (C02) \par
    (Google Assistant returns a list of links. C02 browses the list) \par
    \textit{“It’s not what Ms. L said. I’ll spell the webpage link.”} (C02)\par
   \textit{ “You’re not supposed to spell the webpage. Look, it’s huge. Google Assistant won’t find it.”}(P02)\par
\end{quote}

Teachers also provided troubleshooting tips for retrieving information from CAs. These tips, though not specific to child-CA interactions, supported students relying on CAs for homework. For example, a teacher's instruction to search for images before refining a query illustrated proactive guidance to optimize search results: \textit{“Look, here’s what Ms. R said to do. ‘Hey Google, show me pics about the book A Chat About Shoes. See, I’ll choose this one to make the model.”} (C06).

Across these scenarios, MKOs calibrated the balance between independence and support—modeling prompt-formulation, redirecting unproductive trial-and-error, and affirming correct steps, so that children gradually assumed greater control over their CA-mediated problem solving.

\subsubsection{Functional purposes and workarounds in child-CA interaction}
Children engaged with CAs in home, school, and public settings (e.g., malls, parks) to support three primary functions: school, entertainment, and task automation. Although our analysis centred on CA-assisted homework, we observed that knowledge discovery, entertainment and automation tasks frequently intertwined with academic work, serving both to sustain motivation and to offer brief, non-disruptive breaks during study sessions.

The following sections will detail the activities and strategies children used with CAs for school, entertainment, and task automation.

\begin{figure}
    \centering
    \includegraphics[width=0.95\linewidth]{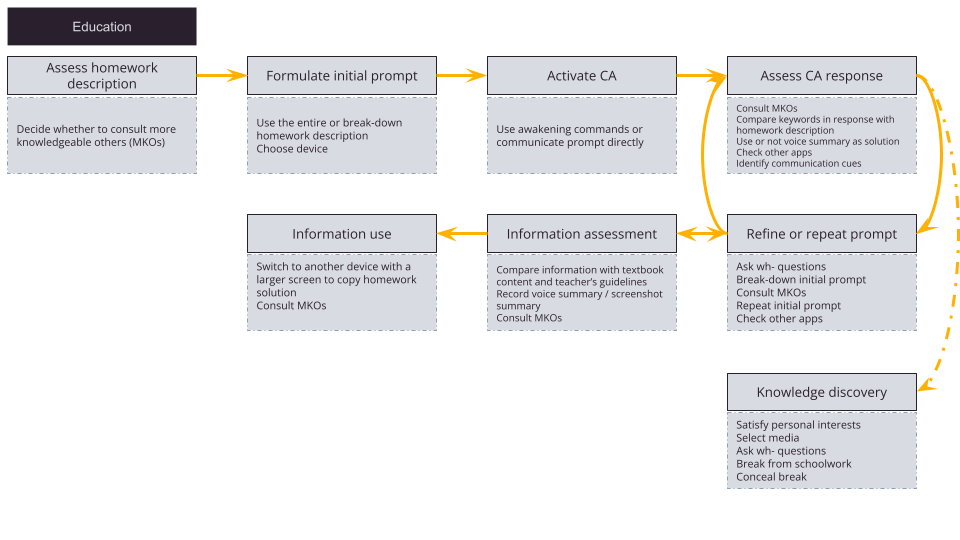}
    \caption{Key tasks in CA‐supported homework: formulating prompts, retrieving information, and applying results. The diagram highlights iteration—query refinement, consulting more knowledgeable others, and switching sources—and highlights how CAs and social support jointly facilitate independent learning.}
    \Description{This figure illustrates key tasks such as prompt formulation, information retrieval, and the utilization of retrieved data for homework completion. The diagram also shows the iterative nature of these interactions, including the decision points where children may seek help from more knowledgeable others (e.g., parents, teachers, peers), refine their queries, or switch to alternative information sources. The figure underscores the role of CAs in facilitating independent learning and the support mechanisms children leverage to enhance their educational outcomes.}
    \label{fig:educationfunction}
\end{figure}

\paragraph{\textbf{Education. }} Children used CAs to research homework by: (1) reading and reflecting on the assignment, (2) formulating a voice query, (3) assessing the CA’s response, and (4) consulting more knowledgeable others (MKOs) as needed to refine prompts or interpret results (Figure \ref{fig:educationfunction}).

\begin{itemize}
    \item Decisions and challenges: In prompt formulation, children chose between using the full homework wording or asking an MKO for clarity. If a CA reply was unclear, they repeated prompts slowly, broke queries into smaller parts, or asked parents/teachers for alternative phrasing. When voice summaries felt incomplete, they switched to browsing results or compared them against textbooks. Alternatively, they switched to a larger screen (e.g., desktop computer, laptop, tablet) to improve accessibility while copying the selected homework solution.
    \item Workarounds: To avoid overwhelm, children assumed the top CA result was best (\textit{"Google Assistant “always gives the best answer”}, C06) and, if necessary, jumped to specialized apps (e.g., Brainly, AprendiZap) to decode terminology or retrieve missing details.
\end{itemize}

\begin{figure}
    \centering
    \includegraphics[width=0.95\linewidth]{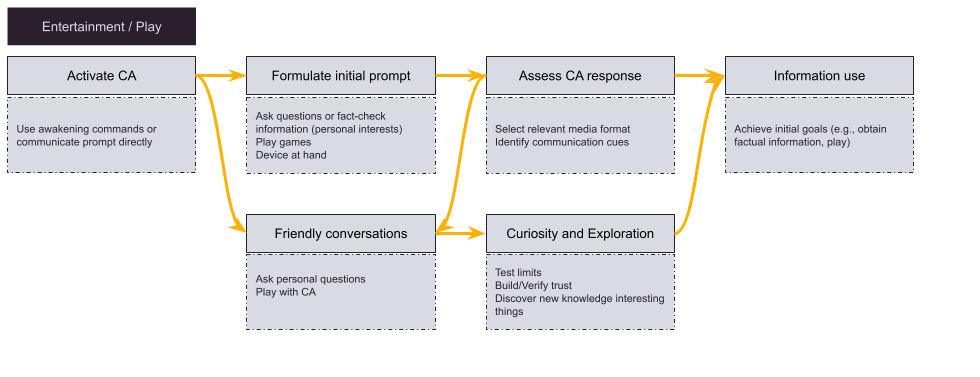}
    \caption{Children’s CA leisure activities—jokes, games, TV queries, and interest exploration, showing iterative command refinement and clarifications that support creativity, curiosity, and engagement.}
    \Description{The figure showcases key tasks such as requesting jokes, playing games, asking about TV shows, and exploring personal interests. The diagram also highlights the iterative interactions where children may refine their commands, seek clarifications, or repeat prompts to achieve the desired outcomes. Additionally, it illustrates how these interactions foster creativity, curiosity, and leisure, providing children with a fun and engaging experience through their use of CAs.}
    \label{fig:entertainmentplayfunction}
\end{figure}

\paragraph{\textbf{Entertainment. }} CAs served as spontaneous companions for facts, jokes, games, or media queries (Figure \ref{fig:entertainmentplayfunction}).

Children sometimes encountered issues, such as accidentally closing tabs and losing access to the previously displayed information, leading to frustration when they could not easily retrieve the same results:
\begin{quote}
    \textit{“I want to show that page we found”} (C02)\par
    \textit{“Ask the same thing and you’ll find it”} (P02)\par
    \textit{“Hey Google, I want to learn how to make cornstarch quicksand.”} (C02)
    (Google Assistant returns a list of links)
    \textit{“These things aren't the things we found last time (C02 points to the phone screen). I lost it.”} (C02)\par
\end{quote}

\begin{itemize}
    \item Decisions and challenges: Children frequently paused their homework to pursue curiosity-driven queries—“sometimes I record what (Google Assistant) is reading. But I need to be super quick.” (C01)—then returned to study. When voice summaries were absent during gaming or social media, they diverted to other sources (e.g., YouTube, Brainly) to fill gaps. For example, C08 attempted \textit{"Play Mia and Me theme song. Hm, it didn't work. Open YouTube. Ooo-pennn OOO Tooobeh", revealing how pronunciation difficulties forced repeated reformulations.}
    \item Workarounds:
    \begin{enumerate}
        \item Four children (C01, C03, C04, C06) recorded CA audio on their phones to avoid losing relevant information, especially when they could not complete their homework in one session (\textit{“sometimes I record what (Google Assistant) is reading. But I need to be super quick.” C01).}
        \item Spontaneous play queries also inspired workarounds. After asking Siri about a new season of their favourite show—\textit{“I want to know if there’ll be another season for Mia and me and when this season will be released”—C08 added, \textit{“that’d be cool if Siri could give that information right away, so I wouldn’t need to say all these things to her.”} When Siri failed, C08 switched mid-query to YouTube, repeating “Open YouTube” until the CA recognized the command.}
        \item Similarly, when nonsensical CA replies confused children—\textit{“I like her. She knows everything.”} (C09) contrasted with awkward jokes—\textit{“My pet is a banana”} (Google Assistant responding to C09)—children simply re-engaged the CA, believing persistence would yield better results: \textit{“sometimes it’s a lot of work to just get the CA giving the right answer.”} (C07). These workarounds—audio recording, rapid prompt switching, and persistent rephrasing—allowed children to reclaim control when CAs did not meet their playful expectations.
    \end{enumerate}
    \ 
\end{itemize}

\paragraph{Task Automation. } Children encountered difficulties in formulating precise commands for CAs, particularly when setting up tasks such as reminders or adjusting device settings like language preferences, \textit{“I changed my nickname a while ago, but I don’t know how to do this anymore.”} (C05). To address these challenges, children sought direct guidance from the CA, highlighting a learning process wherein they acquired new skills through interactive assistance, like C05 \textit{“Hey CA, how do I change my nickname?”}

CAs also assisted with tasks like opening apps (e.g., Netflix, YouTube), checking weather forecasts, and playing videos or music, \textit{“I’ll travel to Raposo in two months, so I asked Siri to tell me what the weather will be like.”} (C04). Children said that automating app access \textit{“made things a lot easier, because I didn’t need to open those apps”} (C06), enhancing efficiency in accessing entertainment and informational content. 

\begin{figure}
    \centering
    \includegraphics[width=0.95\linewidth]{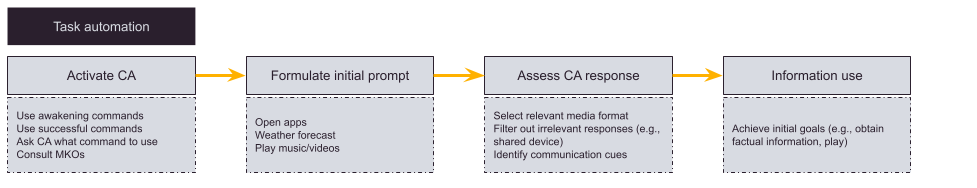}
    \caption{Activities and processes children use CAs for task automation—setting reminders, changing settings, opening apps, checking weather, and playing media—highlighting how they refine commands (e.g., repeats, prompt edits) and rely on more knowledgeable others to troubleshoot and streamline routines.}
    \Description{This figure illustrates the activities and processes children engaged in while using CAs for task automation. Key tasks include setting reminders, adjusting device settings, opening apps, checking weather forecasts, and playing music or videos. The diagram emphasizes how children navigated communication challenges, such as repeating commands and modifying prompts, to effectively automate routine tasks. It also highlights the support from more knowledgeable others  in refining commands and troubleshooting issues, demonstrating how CAs streamline and simplify everyday activities for children.}
    \label{fig:taskautomationfunction}
\end{figure}

\begin{itemize}
    \item Decisions and challenges: Modifying commands based on feedback received from MKOs or even the CA, or adjusting prompts led to challenges because children faced difficulties in such commands. Another challenge/decision-point resided in shared family profiles which skewed search results towards adult-oriented content, like what C08 experienced: \textit{“Who was looking for Kauai clothes? Why is it showing here? Am I getting a birthday gift?”}
    \item Workarounds: Strategies like speaking slowly, breaking down prompts into simpler parts, or seeking external assistance helped overcome communication barriers with CAs. As for profile differentiation, there was no workaround.
\end{itemize}

\subsubsection{Influence of goals, constraints and priorities in child-CA interactions in home contexts}
Children turned to CAs at home both to discover new knowledge and to unwind. However, these goals often clashed with practical and social limits.
\paragraph{Goals and constraints. } Children wanted quick, accurate answers for homework and spontaneous fun, yet their CA use was bounded by unstable internet, shared devices, and mismatched expectations. Limited data plans forced children to adapt their queries—\textit{“My mom said that I can’t watch videos all the time because I won’t have enough data to look up things for school. So, I have to say ‘show me pictures and text of this and that.’”} (C03). Slow or outdated hardware added friction: \textit{“I asked her (Google Assistant) ages ago and still nothing…"} (C08). P08 stated that: \textit{The tablet is very slow, but there’s only so much I can do to help him.”} Another issue with shared devices was the loss of previously stored interaction data. On top of technical barriers, teachers worried that CAs’ ready-made summaries discourage critical thinking: \textit{“My students say whatever words without thinking about what they want to find. Then, the CA gives the answer, the perfect summary, leading them to believe that the summary has the solution. When I see those fancy words, I’ll confront students. Most of the time, they’ll admit that they don’t know what they copied.”} (T01). Finally, because CAs were built for general automation—not schooling—children sometimes struggled to find developmentally appropriate content. As T02 put it: \textit{“should present information like a puzzle, requiring students to figure out what’s missing and how to organize the information to make sense.”}.

\paragraph{Priorities. } Children's priorities comprised experiencing serendipitous knowledge discovery, and switching between academic and leisure activities without upsetting their parents. They delighted in “rabbit-hole” moments—\textit{“She goes down a rabbit hole, a knowledge discovery.”} (P04)—yet also needed quick breaks without drawing parental concerns. For instance:

\begin{quote}
    \textit{“Hey Google, play the sound a bison makes.”} (C09)\par
   \textit{ “Here’s the sound a bison makes.”} (Google Assistant)\par
    (C09 giggles).\par
    \textit{“Hey C09, what are you doing? Get back to your homework.”} (P09)\par
\end{quote}

\subsubsection{Translating child-CA interaction patterns into conversation-trees}\label{translating}
We identified three frequent knowledge scaffolding approaches that emerged from situations in which children needed help to overcome barriers in their interactions with CAs.  These knowledge scaffolding approaches emerged in three scenarios: (1) direct help, (2) intervention/indirect help, and (3) help hesitation.
Below we describe these knowledge scaffolding approaches using quotes from our study that illustrate best each level emerging in each of these three scenarios. 
\paragraph{Direct help}
In the \textit{Direct Help} knowledge scaffolding, children sought immediate assistance from more knowledgeable individuals (e.g., parents and peers) as soon as they encountered difficulty with their homework.
\begin{itemize}
    \item Initial Confusion: Child reads the homework description and immediately feels lost.
    \begin{itemize}
        \item Child: \textit{"I don’t know what to do. This doesn’t make sense."} (C05)
    \end{itemize}
    \item Seeking Immediate Assistance: Child asks a parent or peer for help.
    \begin{itemize}
        \item Child: \textit{"How should I ask this to Google Assistant?"} (C09)
        \item Parent: Provides direct instructions or the answer. \textit{“Let’s check the homework description for keywords. Look, you need to pick one of these options: eat, live or life expectancy. Which one would you like to pick?”} (P09)
        \item Child: \textit{“I want to know what toucans eat.”} (C09)
    \end{itemize}
    \item Using Provided Input: Child uses the provided input to interact with the CA.
    \begin{itemize}
        \item Child: \textit{"Hey Google, what does a toucan eat?"} (Using the prompt provided by the parent/peer) (C09)
        \item Child: \textit{“I’ll check what Ms. Y said that I needed to do.”} (Using the guidelines provided by the teacher) (C09)
    \end{itemize}
    \item Evaluating CA Response: Child checks the CA’s response.
    \begin{itemize}
        \item Child: \textit{"Is this correct, mom? It looks correct, but it’s very long."} (C01)
        \item Parent: \textit{"You should read everything and make sense of what you read. Then, you’ll understand what information you should use to answer your homework."} (P01)
    \end{itemize}
    \item Outcome: the child completed the task with varying levels of understanding, depending on whether they were encouraged to think critically or just use the provided answers.
\end{itemize}

Without attempting to fully understand the homework requirements, children relied on parents or peers to provide direct answers or specific phrases to use with the CA. This approach often involved straightforward requests like, \textit{“I don’t know what to do. Help me, (P05). This doesn’t make sense”} (C05) and led to direct intervention where the helper (i.e., parent) offered the exact wording or solution needed. The child's reliance on direct input from others indicated children’s need to decode the homework instructions, formulate effective prompts to CAs and find suitable information sources.

\paragraph{Indirect help}
The \textit{Indirect Help} knowledge scaffolding emerged when children unknowingly faced problems during their interactions with the CA. These scenarios were marked by a trial-and-error method where children repeatedly used the same prompts, expecting different outcomes.
\begin{itemize}
    \item Stuck in Trial and Error: Child encountered a barrier to solving a problem and insisted on using the same prompt.
    \begin{itemize}
        \item Child: \textit{"Hey Google, what are \textit{cabaças}?"} (C04 speaking slowly in her second attempt) (C04)
        \item CA: "I don’t understand what you’re saying."
        \item Parental Intervention: Parent noticed the child’s repeated unsuccessful attempts and intervened.
        \item Parent: "Explain to me what you’re doing." (P04)
        \item Child: "Here, I’ll do it again so you can see." (C04)
        \item Parent: "Let’s come back to the textbook. What else is there?" (P04)
    \end{itemize}
    \item Guidance to Refocus: Parent redirected the child to another approach. Children also referred to what their teachers recommended when the homework was assigned in class.
    \begin{itemize}
        \item Parent: \textit{"Let’s try this time ‘Brazilian Indigenous musical instruments made with dried gourd’."} (P04)
        \item Teacher: \textit{“I usually give them some keywords they can use to search for homework information.”}(T01 recommended when the homework was assigned in class) (T01).
        \item Child: Repeated the new prompt or used the keywords the teacher provided, and receives relevant information. \textit{“Hey Siri, show me some Brazilian Indigenous musical instruments made with dried gourd.”} (C04)
    \end{itemize}
    \item Distraction Management: Parent helped child stay focused when distracted by interesting but irrelevant information.
    \begin{itemize}
        \item Parent: \textit{"Have you found the information about African-Brazilian religions?"} (P01)
        \item Child: \textit{"Look, mom! This is so interesting. Did you know that they blended Catholic and African saints?"} (C01)
        \item Parent: \textit{"Will all of this be in your homework?"} (P01)
        \item Child: \textit{"No, I just found it interesting."} (C01)
        \item Parent: \textit{"Look it up after you finish your homework."} (P01)
    \end{itemize}
    \item Outcome: Child completed the homework with a better understanding and maintained interest in additional learning.
\end{itemize}

Parents observed the child's struggle and intervened by asking probing questions to identify the obstacles, while teachers provided guidelines to help children retrieve information using CAs. The intervention helped redirect the child’s efforts and introduced alternative approaches to find the information needed. This conversation flow emphasized the role of adult guidance in helping children recognize ineffective strategies and encouraging a more focused and methodical approach to their homework activities.

\paragraph{Help hesitation}
The \textit{Help Hesitation} knowledge scaffolding described scenarios where children felt reluctant to ask for help, often due to a perceived simplicity of the task or previous experiences of minimal encouragement. These children recognized that they needed assistance but hesitated to seek it, fearing judgment or thinking they should be able to solve the problem independently.
\begin{itemize}
    \item Self-Doubt: Child felt embarrassed or hesitant to ask for help, thinking the task should be easy.
    \begin{itemize}
        \item Child: \textit{"I feel so stupid for not seeing that the answer was right there. I could’ve asked my mom to help me, but she’d say that it was so easy."} (C06)
        \item Mother: \textit{"Sometimes, he doesn’t want to make the effort to solve the homework task, even the easiest ones."} (P06)
    \end{itemize}
    \item Struggling with Prompts: Child struggled with formulating effective prompts for the CA.
    \begin{itemize}
        \item Child: \textit{"I want to find things about tropical forests. Wow, it’s always returning these things. It’s not what I want."} (C07)
    \end{itemize}
    \item Repeated Attempts: Child kept trying similar prompts with little success.
    \begin{itemize}
        \item Child: \textit{"I feel so stupid. It always worked other times."} (C05)
    \end{itemize}
    \item Lack of Intervention: Parent did not intervene, leaving the child to navigate their struggle alone.
    \begin{itemize}
        \item Child: \textit{“I’ll try it again. I’ll say ‘what minerals have purple colour?’ (C03 looks at her mom)."} (C03)
        \item Mother: \textit{“Keep trying.”} (P03)
    \end{itemize}
    \item Outcome: The child needed reassurance and guidance to understand what parts of their interactions were or weren’t working, which they often did not receive, leading to frustration and halted progress.
\end{itemize}

Children’s interactions with the CA became fraught with self-doubt and repeated, ineffective attempts to retrieve the desired information. The lack of parental intervention in these cases further compounded the child's frustration and uncertainty. This conversation flow highlighted the need for reassurance and guidance to help children understand their interaction challenges and build confidence in their problem-solving abilities.

\section{Study 2—Conversation-tree evaluation}
Study 2 builds on the findings in Study 1 of how parents scaffold homework discussions, implicitly using strategies like Questioning the Author (QtA) and Vygotsky’s Zone of Proximal Development (ZPD). These strategies helped children not only find answers but also think through problems and justify their reasoning. In the following sub-sections, we discuss the design and implementation of a conversation-tree recipe based upon structured-prompting, the study design and the results.
\subsection{Conversation-tree recipe}
In Study 1, we identified core turn-by-turn interaction patterns that support cognitive scaffolding in parent-child dialogues (see sub-section \ref{translating}). Based on these patterns, we designed and tested a conversation-tree recipe using a large language model (LLM). We opted to design and implement the conversation-tree recipe using LLMs due to its capabilities to generate culturally-driven interactions and dynamic conversations that adapt to users' contexts. The conversation‑tree “recipe” can be applied to constrain LLMs across five dimensions: system (i.e. LLM), mode, learning customization, learning assessment and game generation, so that each child‑CA interaction remains developmentally appropriate, pedagogically scaffolded, and engaging.

\subsubsection{System boundaries.} At the highest level, a \texttt{system prompt} enforces the conversation-tree recipe:
\begin{itemize}
    \item \textit{Start by requesting the child's grade level and desired mode (school, discovery or entertainment).}
    \item \textit{Adjust all subsequent turns around that grade and mode, persistently carrying context.}
    \item \textit{Knowledge scaffolding by encouraging critical thinking and avoiding direct answers; instead, the system asks questions that probe the child's current understanding and knowledge level.}
\end{itemize}

In \texttt{school} and \texttt{discovery} modes, the system additionally asks the child to self‑report their \texttt{knowledge level} (“little,” “some,” or “a lot”) before scaffolding. In \texttt{entertainment mode}, after grade entry, the \texttt{system} asks if the child would like to play a game, and, if confirmed, generates a grade‑appropriate puzzle or riddle.

\subsubsection{Mode boundaries.} We assign the LLM a distinct \texttt{role} and \texttt{tone profile} per mode, instantiated with grade‑specific descriptions derived from Piaget’s developmental stages \cite{piaget_psychology_1973} and Flesch‑Kincaid readability targets \cite{KincaidJP1975DoNR}. Piaget’s four stages chart how thinking evolves: first, sensorimotor infants (0-2 years) learn through action and gain object permanence; next, preoperational children (2-7 years) use symbols but lack logic; then, concrete operational children (7-11 years) reason about real events; and finally, formal operational youths (11+ years) handle abstract, hypothetical thought. Flesch–Kincaid formulas classify text difficulty using average sentence length and syllables per word: the Reading Ease yields a 0–100 ease score, while the Grade Level indicates the grade needed to understand the text.

We expand below the description crafted for each mode in the recipe:
\begin{itemize}
    \item \textbf{School—Tutor role:} \texttt{The tutor will use a simple and concrete language that is based on student's everyday experiences when explaining concepts. The tutor will focus on foundational skills like basic problem-solving, counting, and identifying patterns} (\texttt{tone profile} for grade 1).
    \item \textbf{Discovery—Guide role:} \texttt{The guide is interacting with a Grade 6 student, challenging them to explore problems using abstract reasoning. The guide guides them in using logical steps and hypothesis testing} (\texttt{tone profile} for grade 6).
    \item \textbf{Entertainment—Friend role:} \texttt{The friend presents advanced challenges that require systematic reasoning, creative solutions, and deep analysis} (\texttt{tone profile} for grade 11).
\end{itemize}

Having a \texttt{profile tone} for each grade in all modes provides an ongoing check so that vocabulary, sentence complexity and conversational tone align with the child's cognitive and literacy abilities.

\subsubsection{Learning customization.} Once grade, mode, and task (\texttt{school}) or topic (\texttt{discovery}) are defined, the \texttt{system}:
\begin{itemize}
    \item Assesses topical knowledge via the child's self-rating (e.g., little, some or a lot).
    \item Adjusts knowledge scaffolding by using a list of examples drawn from Study 1 as in-prompt references.
    \item Monitors for fallback cues (e.g., "I don't understand") and, if detected, reduces complexity and switches to mode foundational knowledge scaffolding.
\end{itemize}

The strategies above create customization boundaries, improving dynamic support and adjustment just-in-time.

\subsubsection{Learning assessment.} In both \texttt{school} and \texttt{discovery} modes, we included a \texttt{learning assessment} component to evaluate child's learning and further support critical thinking. After the task/topic completion is indicated, the \texttt{system} quizzes the child. Correct answers earn reinforcement; incorrect answers trigger another scaffolding cycle.

\subsubsection{Game generation. } In the \texttt{entertainment mode}, the \texttt{system} generates riddles, puzzles and brain teasers based on game templates calibrated by grade level. To ensure that the games are fun and remain developmentally appropriate, we also referred to Piaget’s developmental stages \cite{piaget_psychology_1973} and Flesch‑Kincaid readability targets \cite{KincaidJP1975DoNR}.
\begin{itemize}
    \item \textbf{Game template: } \texttt{Create intellectual riddles or puzzles for a Grade 12 student. These riddles or puzzles should require abstract thinking, advanced reasoning, or literary knowledge (Grade 12).}
\end{itemize}

\subsection{Study design}
To evaluate the effects of our conversation‑tree recipe, we conducted a controlled comparison between a GPT 4o‑mini model conditioned by a structured-prompting (i.e., conversation-tree recipe) against the unmodified base model without any prompt conditioning. We selected GPT 4o‑mini both for its ability to support dynamic, culturally‑aware dialogue and because its token‑cost profile made it feasible to run our large test suite. The conversation-tree recipe was implemented as a customized prompt via OpenAI API.

We generated a corpus of simulated “child–CA” exchanges covering all three interaction modes (school, discovery, entertainment) for four grade bands (1, 5, 9, 12). \footnote{The dataset and analysis report generated in this study will be made publicly available upon acceptance of this manuscript.} We opted to focus on these four grades because they exhibit the most significant differences in cognitive and literacy abilities \cite{piaget_psychology_1973, KincaidJP1975DoNR}, which helps illustrate model performance across developmental stages.

\subsubsection{Stimuli generation:}
\begin{itemize}
    \item \textbf{School and discovery:} 5 prompts × 3 subjects (math, science, Brazilian social sciences) × 4 grades=60 prompts per mode. Each prompt was issued under 3 knowledge‑levels and 3 temperatures, yielding 540 test‑cases per mode.
    \item \textbf{Entertainment:}  5 puzzles per grade × 4 grades=20 prompts; each at 3 temperatures, with 5 simulated child replies per prompt (three incorrect and two correct, for 60 cases. This configuration allowed us to assess the model’s hint‑generation and adaptive scaffolding behaviour.
\end{itemize}

\begin{table*}[ht]
  \centering
  \caption{\quad The table presents three illustrative prompt–answer pairs used during the stimulus‑generation process, organized by instructional mode, student grade, and subject area. For each prompt, the corresponding gold‑standard response shows how an ideal answer both models the expected reasoning steps and invites further student reflection.}
  \label{tab:prompt-examples}
  \begin{tabularx}{\textwidth}{ 
      l   
      l   
      c   
      >{\raggedright\arraybackslash}X   
      >{\raggedright\arraybackslash}X   
    }
    \toprule
    Mode & Grade & Subject & Prompt & Gold answer \\
    \midrule
    School  & Grade 1                  & Math & Solve for x: 2x + 3 = 7 &   First, let's isolate the term with x. (...) What do you get then?  \\
    \midrule
    Discovery  & Grade 5                  & Brazilian social sciences   & What is celebrated during June Festival in Brazil? &   June Festival is a big celebration in Brazil! (...) What do you think is special about the traditional foods and dances during this celebration?  \\
    \midrule
    Entertainment  & Grade 12                 & N/A & \textbf{S:} You have two ropes. Each takes exactly one hour to burn, but not at a consistent rate. How can you measure exactly 45 minutes? \textbf{C:} I don't know & Hint: You can light one rope from both ends, and think about when and how to light the second one. \\
    \bottomrule
  \end{tabularx}
\end{table*}

We saved each model reply and computed: (1) Similarity score (cosine similarity to gold scaffolded answer), (2) Readability levels (Flesch-Kincaid formula), (3) Question scaffolding metrics: question count (\texttt{q\_count)}, diversity (\texttt{q\_diversity}), depth (\texttt{q\_depth}), and (4) Latency (response time in seconds).

\subsection{Data analysis}
The goal of study 2 was to answer \textit{RQ6. To what extent do conversation-tree recipes improve knowledge scaffolding, critical thinking and independent problem-solving in CA interactions compared to unstructured, free-form dialogue in LLMs?}. To answer this RQ, we tested the the hypotheses described in Table \ref{tab:analysis-plan}.

\begin{table*}[ht]
  \centering
  \caption{Analysis plan linking each hypothesis about fine‐tuning and temperature effects (H1–H4) to the corresponding statistical test or model used to evaluate grade‐level alignment, question scaffolding, content coherence, and interaction metrics.}
  \label{tab:analysis-plan}
  \begin{tabularx}{\textwidth}{ 
      >{\raggedright\arraybackslash}X   
      >{\raggedright\arraybackslash}X   
    }
    \toprule
    \textbf{Hypothesis} & \textbf{Method}  \\
    \midrule
    \addlinespace[0.5ex]
    \textbf{H1 Grade‐level alignment:} Structured-prompting conditioned outputs will more closely match target readability levels than vanilla outputs.
      & Welch’s ANOVA on similarity score by grade level. \\
    \addlinespace
    \textbf{H2 Knowledge scaffolding:} Structured-prompting conditioned outputs will include more and higher‐quality prompting questions to foster critical thinking.
      & Poisson GLMs for \textit{q\_count} and \textit{q\_diversity}; Mann–Whitney U for \textit{q\_depth} (question‐scaffolding metrics). \\
    \addlinespace
    \multirow{2}{*}{%
      \parbox{\hsize}{%
        \textbf{H3 Recipe effectiveness:} Structured-prompting conditioned outputs will be more coherent, contextually appropriate, and interactive than vanilla outputs.\\[4pt]
        \textbf{H4 Temperature effects:} Higher temperature in the structured-prompting conditioned model will yield content that remains well aligned to grade‐level expectations, more so than in the vanilla model.
      }%
    }
      & Mixed ANOVA (configuration $\times$ temperature), Welch’s t‐test, and multiple regression controlling for readability level, grade level, and temperature. \\
      & \\
      & \\
      & \\
      & \\
    \bottomrule
  \end{tabularx}
\end{table*}

\subsection{Results}
To evaluate our hypotheses about grade‑alignment and scaffolding, we compared structured-prompting versus vanilla GPT 4o‑mini outputs on similarity scores and question‑scaffolding measures. Overall, structured-prompting replies not only matched target readability levels more closely but also generated substantially more—and deeper—questions than the vanilla model.
We organize our quantitative findings around the four hypotheses (H1–H4) concerning grade‑level alignment, knowledge scaffolding, recipe effectiveness, and temperature effects.

\subsubsection{H1: Grade-level alignment}
Levene’s test indicated unequal variances ($W = 15.10$, $p < .001$), and residuals departed from normality (Shapiro–Wilk $W = 0.873$, $p < .001$), so we report Welch’s ANOVA. There was a small but significant effect of grade level on similarity score, Welch’s $F(3,\,1659.8) = 15.16$, $p < .001$, $\eta^2 = 0.015$. 

\begin{figure}
    \centering
    \includegraphics[width=0.8\linewidth]{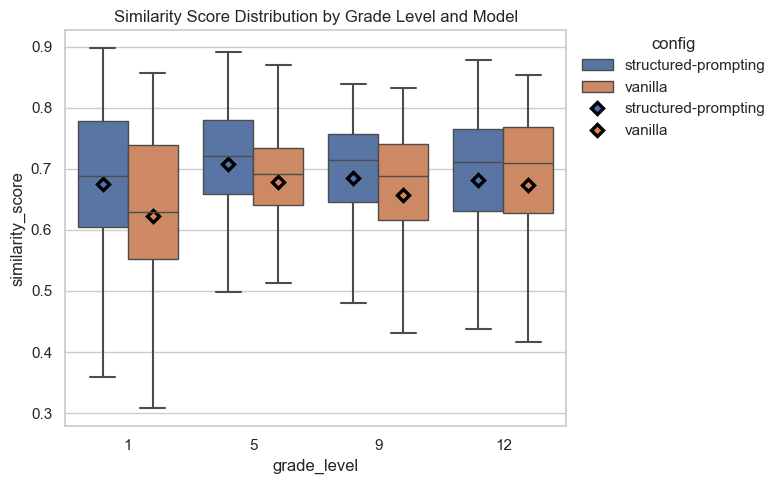}
    \caption{Across grades 1–9, the model conditioned by structured-prompting achieved higher similarity scores than the vanilla model (Welch’s $t = 6.16$, $p < .0001$).}
    \Description{Across grades 1–9, the model conditioned by structured-prompting achieved higher similarity scores than the vanilla model (Welch’s $t = 6.16$, $p < .0001$).}
    \label{fig:enter-label}
\end{figure}

Post‑hoc Tukey comparisons (Table \ref{tab:tukey-similarity-grade}) showed that similarity was higher at grade 1 than at grades 9 and 12 (\$p<.001\$).

\begin{table*}[ht]
\centering
\caption{Tukey’s HSD pairwise comparisons of mean similarity scores across grade levels (family-wise error rate = 0.05).}
\label{tab:tukey-similarity-grade}
\begin{tabularx}{\textwidth}{
    >{\raggedright\arraybackslash}X 
    >{\raggedright\arraybackslash}X 
    S[table-format=1.4]              
    S[table-format=<1.4]             
    S[table-format=1.4]              
    S[table-format=1.4]              
}
\toprule
\textbf{Group 1} & \textbf{Group 2} & \textbf{Mean Diff.} & \textbf{Adj.\ p-value} & \textbf{Lower CI} & \textbf{Upper CI} \\
\midrule
1  & 5  & 0.0441  & <0.001 & 0.0268  & 0.0613  \\
1  & 9  & 0.0219  & 0.0063 & 0.0046  & 0.0392  \\
1  & 12 & 0.0282  & 0.0002 & 0.0109  & 0.0454  \\
5  & 9  & -0.0222 & 0.0054 & -0.0395 & -0.0049 \\
5  & 12 & -0.0159 & 0.0842 & -0.0332 &  0.0014 \\
9  & 12 &  0.0063 & 0.786  & -0.0110 &  0.0236 \\
\bottomrule
\end{tabularx}
\end{table*}

\subsubsection{H2: Knowledge scaffolding}
The structured-prompt recipe yielded markedly richer questioning behaviour across all three metrics (Table \ref{tab:question_scaffolding}).

\begin{table}[ht]
\centering
\caption{Statistical tests of conversation‑tree recipe effects on question‑scaffolding metrics.  Poisson models compare question counts and diversity between structured-prompting vs.\ vanilla; Mann–Whitney compares question depth.}
\label{tab:question_scaffolding}
\begin{tabularx}{\textwidth}{l X c c c}
\toprule
\textbf{Metric} & \textbf{Test (model)} & \textbf{Statistic} & \textbf{p‑value} & \textbf{Effect size} \\
\midrule
Question count & Poisson GLM: \(q_{\mathrm{count}}\sim\) config + mode + grade\_level & \(z=-22.71\) & \(<.001\) & IRR = 0.004 \\
Question depth & Mann–Whitney \(U\) on \(q_{\mathrm{depth}}\) by config & \(U=1\,899\,953.5\) & \(<.001\) & \(r=0.167\) \\
Question diversity & Poisson GLM: \(q_{\mathrm{diversity}}\sim\) config + mode + grade\_level & \(z=-38.38\) & \(<.001\) & IRR = 0.26 \\
\bottomrule
\end{tabularx}
\end{table}

A Poisson generalized linear model on question count revealed that vanilla outputs produced virtually no scaffolded questions compared to the model constrained by structured-prompting (IRR = 0.004, \(z=-22.71\), \(p<.001\)), demonstrating that our recipe reliably increases the quantity of learner‑directed prompts. Likewise, a Mann–Whitney \(U\) test on question depth showed that fine‑tuned replies ask substantially more cognitively demanding questions than vanilla replies (\(U=1\,899\,953.5\), \(p<.001\), rank‑biserial \(r=0.167\)), indicating elevated engagement with higher‑order thinking. Finally, question diversity—measured as the variety of interrogative forms—was also significantly greater under structured prompt condition: Poisson regression produced an IRR of 0.26 for vanilla versus structured-prompting conditioned outputs (\(z=-38.38\), \(p<.001\)), confirming that our conversation‑tree recipe broadens the range of scaffolding prompts. Together, these results provide strong evidence that the structured prompt condition not only increases how many questions the model asks, but also enhances their depth and variety, thereby better supporting critical thinking.

\subsubsection{H3: conversation-tree recipe effectiveness}
A mixed‑design ANOVA revealed a robust main effect of configuration, \(F(1,2996)=37.90\), \(p<.001\), partial \(\eta^2_p = 0.012\); structured-prompting outputs were more coherent than vanilla (Table~\ref{tab:mixed_anova}).

\begin{table}[t]
\caption{Mixed‑design ANOVA for similarity scores (config $\times$ temperature).}
\centering
\small
\begin{tabular}{lrrrrr}
\toprule
Effect & df$_1$ & df$_2$ & $F$ & $p$ & $\eta_p^2$ \\
\midrule
Configuration (config)      & 1 & 2996 & 37.90 & $<.001$ & 0.012 \\
Temperature                 & 1 & 2996 &  0.27 & 0.61    & 0.000 \\
Config $\times$ Temperature & 1 & 2996 &  0.22 & 0.64    & 0.000 \\
\bottomrule
\end{tabular}
\label{tab:mixed_anova}
\end{table}

There was no significant main effect of temperature, \(F(1,\,2996)=0.27\), \(p=0.61\), nor a config\(\times\)temperature interaction, \(F(1,\,2996)=0.22\), \(p=0.64\). Welch’s \(t\)-test comparing overall similarity by configuration corroborated this: \(t(\approx2980)=6.16\), \(p<0.001\), Cohen’s \(d=0.225\), 95\% CI \([0.150,0.300]\) (Table~\ref{tab:welch_t}).

\begin{table}[t]
\caption{Welch’s $t$‑test comparing similarity by configuration.}
\centering
\small
\begin{tabular}{lrrrrr}
\toprule
Comparison & $t$ & df & $p$ & Cohen’s $d$ & 95\% CI \\
\midrule
Structured prompt vs.\ Vanilla & 6.16 & $\approx$ 2980 & $<.001$ & 0.225 & [0.150, 0.300] \\
\bottomrule
\end{tabular}
\label{tab:welch_t}
\end{table}

\subsubsection{H4: Temperature effects}
No main effect of temperature, $F(1,2996)=0.27$, $p=0.61$, nor config\,$\times$\,temperature interaction, $F(1,2996)=0.22$, $p=0.64$ (Table \ref{tab:anova_temp}).

\begin{table}[t]
\caption{Mixed‐design ANOVA: temperature effects on similarity.}
\centering
\small
\begin{tabular}{lrrrrr}
\toprule
Effect & df$_1$ & df$_2$ & $F$ & $p$ & $\eta_p^2$ \\
\midrule
Temperature              & 1 & 2996 & 0.27 & 0.61 & 0.000 \\
Config\,$\times$\,Temperature & 1 & 2996 & 0.22 & 0.64 & 0.000 \\
\bottomrule
\end{tabular}
\label{tab:anova_temp}
\end{table}

Multiple regression likewise showed a small negative temperature effect, $\beta=-0.012$, $p=0.027$, but no practical benefit to higher temperature (Table \ref{tab:reg_temp}).
\begin{table}[t]
\caption{Regression coefficient for temperature (predicting similarity).}
\centering
\small
\begin{tabular}{lccccc}
\toprule
Predictor    & $\beta$ & SE   & $t$    & $p$   & 95\% CI \\
\midrule
Temperature  & –0.012  & 0.005 & –2.206 & 0.027 & [–0.022, –0.001] \\
\bottomrule
\end{tabular}
\label{tab:reg_temp}
\end{table}

\section{Discussion}
Our two studies form a coherent pipeline—from rich, field‑based discovery to large‑scale quantitative validation—showing how parent‑style scaffolding strategies can be encoded as LLM “recipes” to improve children’s learning dialogues. In Study 1, seven weeks of video‑elicitation, interviews, and Cognitive Work Analysis with Brazilian 9–11 year olds (and their parents and teachers) revealed three core CA functions—School, Discovery, and Entertainment—each driven by distinct turn‑taking cues, question strategies, and interpersonal supports. Those empirically grounded patterns directly shaped our conversation‑tree recipe. In Study 2, we tested a GPT 4o‑mini conditioned by our structured prompt on 1,200 simulated child–CA exchanges and compared recipe‑structured against vanilla (i.e., free form) outputs. The recipe model constrained by structured-prompting delivered significant, practical gains in scaffolded‑question count, depth, and diversity; in grade‑level readability alignment; and in overall coherence—while sampling temperature had no appreciable effect. By connecting qualitative mappings of real‑world child–CA workflows with rigorous LLM evaluation, we validate a culturally grounded, script‑light framework for dynamic knowledge scaffolding in educational agents.

In the sections that follow, we relate these findings back to prior work and draw out design implications. We organize our discussion around the six research questions posed across Study 1 (RQ1–RQ5) and Study 2 (RQ6).

\subsection{RQ1: How do children structure their information-processing flows when interacting with conversational agents?}
Children’s information-processing flows revealed both ideal and suboptimal problem-solving strategies. Ideal flows, such as revisiting prior successful prompts, echo earlier findings on children reusing effective queries \cite{xu_are_2021, cagiltay_my_2023}. Suboptimal flows surfaced when children hit barriers: they “refound” and replayed previously heard voice summaries or broke down prompts piece by piece to coax the CA into understanding them. These workarounds demonstrate children’s adaptability when familiar obstacles recur. Moreover, information-processing flows often overlapped across functions: education, discovery, entertainment, and task automation occurred within the same session, suggesting that children see CAs as multifaceted companions that provide summaries, hints, and playful diversions.

Attributing human characteristics to CAs reduced interaction barriers, especially for young children \cite{garg_conversational_2020}, but it also led them to believe that CAs were almost human or manipulated by humans, similar to the concept in \textit{The Wizard of Oz}. CAs mimicked conversational exchanges between humans, causing children to develop a theory of mind (ToM) similar to what they construct of their peers. While it was not reported that CAs could have feelings, children understood these agents to have thoughts and perspectives akin to those in human social interactions. This misconception prevented children from fully leveraging CA capabilities. However, this could not be considered the sole reason for children's frustration \cite{marchetti_theory_2018}, as many continued interacting with CAs even after these agents failed to "understand" and respond as expected. Educating children about how CAs function can improve their interactions because children adjust their mental models after acquiring novel knowledge, confirming previous research \cite{williams_is_2019, ward_my_2011, van_brummelen_what_2023}.

\subsubsection{Implications for design}
\begin{itemize}
    \item \textbf{Support replay \& bookmarks.} Offer built-in “rewind” or “bookmark” commands so children can easily return to earlier voice summaries or search results without re-formulating complex prompts. Native replay/bookmark features reduce repetition overhead and let children focus on comprehension rather than mechanical retrieval. 
    \item \textbf{Expose simple CA mechanics.} Briefly explain (“voice tip”) why and when the CA listened or timed out, helping calibrate children’s mental models and reduce frustration. Children develop inaccurate mental models when CA failures seem arbitrary \cite{hiniker_can_2021, lovato_hey_2019}. Lightweight transparency helps calibrate expectations, decreases blame on the child, and supports more effective subsequent queries.
\end{itemize}

\subsection{RQ2: What roles do more knowledgeable others  play in children’s CA-supported problem solving?}
While homework is intended to foster independent learning, family members often provide support for children to complete these activities \cite{wood_role_1976}. Parents, teachers, and peers (More Knowledgeable Others—MKOs) routinely framed problems, suggested keywords, and validated CA outputs. For example, parents prompted children to “read the homework description, then tell me what you understand from it,” scaffolding initial comprehension before any CA query. Teachers provided checklists and sample search phrases, allowing children to refine their voice prompts. When CAs misinterpreted prompts, children turned to MKOs to rephrase queries or confirm whether the CA’s answers were academically sound.

Additionally, CAs started from a standard point each time, regardless of the level and timing of parental support. This finding aligns with previous research \cite{garg_conversational_2020, lovato_hey_2019}, which revealed that parents wanted control over when and how CAs would assist their children. Parents could benefit from configuring the support CAs provide (e.g., direct help, indirect help, and help hesitation) based on the activity type and their availability. While a CA help setup was suggested in prior research \cite{ho_its_2024}, our suggestion expands on this by incorporating children's willingness to seek help from MKOs, which can be assessed based on parents' understanding of their children's behaviours.

Nevertheless, our study also found that children sometimes disagreed with their parent's support, especially when they had a clear idea of how to formulate their prompts. Children assessed whether the proposed strategies aligned with their prior knowledge, experience and teacher recommendations, particularly when teachers provided keywords or sentences to use as prompts. This provides novel insight into child-CA interactions, challenging the common perspective that children are always willing to accept the solutions and strategies proposed by trusted adults (e.g., parents and teachers) \cite{ho_its_2024}.

Prior research emphasized that CAs should support academic collaborations between children and family members \cite{garg_conversational_2020, xu_rosita_2023}. In our study, parents reviewed information on the homework topic using the CA. Despite sharing profiles, parents' interactions and input were not stored or recalled when children used the CA. While parents did not explicitly mention that this feature would be helpful, it could assist them select and curate information sources for academic and entertainment/play activities. This would support children by providing validated information from their parents, while still allowing children to decide which sources to use, think critically about concepts, and solve problems independently.

Attributing human traits to CAs led children to expect Theory of Mind (ToM)-style understanding, aligned with previous studies \cite{marchetti_theory_2018}. When ToM failed, MKOs stepped in to interpret CA responses and bridge the gap.  This dynamic highlights a distributed problem-solving system in which CAs, children, and MKOs each contribute distinct competencies.

\subsubsection{Implications for design}
\begin{itemize}
    \item \textbf{Teacher-curated keyword sets.} Allow teachers to upload sets of class-specific keywords or phrasing templates that the CA can surface as guided suggestions when it detects a homework context. Surfacing these vetted templates reduces trial-and-error, strengthens alignment with classroom expectations, and builds children’s confidence in independent CA use. 
    \item \textbf{Explain-to-parent logs.} Generate a simple log of CA interactions that parents can review, so they can more effectively coach children on prompt formulation and validate content. A transparent log enables caregivers to pinpoint misunderstandings, validate content accuracy, and reinforce learning through targeted guidance \cite{garg_conversational_2020, lovato_hey_2019, han_teachers_2024}.
\end{itemize}

\subsection{RQ3: For what functional purposes do children use CAs, and what workarounds do they develop when CAs fall short?}
Children used CAs concurrently for four core functions—education, discovery, entertainment, and task automation—even within a single session. They began with homework queries, detoured into curiosity-driven “rabbit holes,” and then used CAs to set reminders or control media. When CAs fell short (e.g. mispronunciation issues, lost tabs), children adopted workarounds such as (1) recording CA audio on their phones to capture fleeting voice summaries, (2) switching mid-query to specialized apps (e.g. Brainly, YouTube) when the CA’s answer was unsatisfactory, and (3) persistently rephrasing commands until the CA recognized them. Developmental psychology research shows that children often engage in multiple activities simultaneously, especially in informal learning environments like the home \cite{vygotsky_mind_1978, wood_role_1976}. However, prior research on child-CA interactions did not always elicit this concurrent engagement across education, discovery, entertainment and task automation \cite{garg_last_2022, xu_rosita_2023, druga_hey_2017}.

Both children and parents reported frequent knowledge journeys where children sought information to satisfy their curiosity, highlighting a novel direction that CAs can bridge formal and informal learning. Although commercial CAs are primarily designed as digital assistants, children also use them for academic purposes \cite{druga_family_2022, garg_conversational_2020, du_alexa_2021}. As children explore their favourite topics, hobbies, and games, these informal learning experiences could connect with formal learning activities. For instance, CAs could link homework topics with children's personal interests, encouraging them to see the connections. Although embedding serendipitous knowledge discoveries into CA design is challenging \cite{lee_dapie_2023, kanda_two-month_2007, ward_my_2011}, it can help map patterns in question-answer interactions.

\subsubsection{Implications for design}
\begin{itemize}
    \item \textbf{Persistent context buffers.} Maintain a short history of the last \textit{N} voice interactions, accessible on demand (“What did you just say?”). Children recorded CA audio externally to recapture lost summaries, introducing extra cognitive load and disrupting their flow. An internal buffer preserves context, reduces reliance on ad-hoc workarounds, and lets them focus on learning rather than on managing recordings.
    \item \textbf{Hand-off to apps.} When a query would be better served by a specialized service (e.g. video tutorial), the CA can proactively suggest and link to that app, preserving context. Our participants abandoned CAs mid-query for Brainly or YouTube when answers fell short, losing prior context and needing to start over. Automated, context-preserving hand-offs eliminate this break in continuity and support children’s multitasking across tools. 
    \item \textbf{Speech adaptation.} Integrate child-tuned speech recognition models that adapt to mispronunciations, non-native accents, and variable pacing. Misrecognition of children’s speech forces repeated rephrasing, frustrating young users and breaking engagement \cite{cheng_why_2018, garg_conversational_2020, druga_hey_2017}. Tailored models reduce recognition errors, minimize trial-and-error, and let children concentrate on problem solving rather than on “speaking correctly.”
\end{itemize}

\subsection{RQ4: What goals, constraints, and priorities influence children’s CA interactions in home contexts?}
Our findings reveal that in home settings children balance two often‐competing goals: completion of school‐related tasks and open‐ended exploration driven by curiosity or play. Prior work has shown that children typically use smart speakers for task automation and screen‐based agents for information seeking, but has not characterized how they fluidly switch between academic and leisure activities within a single session \cite{garg_conversational_2020, lovato_siri_2015, sciuto_hey_2018}. We extend this literature by demonstrating that children expect a single CA to support rapid “homework mode” queries and then pivot into “entertainment mode” without reinitializing context or losing progress.

At the same time, we confirm device‐level constraints documented in prior studies. Audio‐only interfaces impede engagement, while unstable internet and low‐performance shared devices limit multimedia use and force children to rephrase queries for text‐only responses \cite{xu_exploring_2020, le_skillbot_2022}. Crucially, our data show that these environmental constraints become primary drivers of interaction strategies: children explicitly request “show me text, not video” under poor bandwidth, and adopt “data‐frugal” workarounds that any robust home CA must anticipate.

We also surfaced a richer portrait of stakeholder priorities. Consistent with earlier reports of parental concerns over content appropriateness \cite{garg_conversational_2020, le_skillbot_2022}, caregivers in our study stressed accuracy, age‐appropriateness, and preserving critical thinking. Novel to our work, children themselves prioritized maintaining parental approval—strategically interleaving brief entertainment “breaks” within homework sessions to avoid reprimand—while still pursuing unstructured “rabbit‐hole” learning. This interplay of child and parent priorities highlights the need for CAs to negotiate trust, autonomy, and oversight dynamically.

By integrating these elements via a Cognitive Work Analysis framework \cite{rasmussen_taxonomy_1990}, we provided the first holistic model of how home‐context goals, resource constraints, and social priorities co‐shape children’s CA use. Whereas prior work catalogued individual barriers or uses in isolation \cite{lovato_hey_2019, ho_its_2024}, our study revealed their dynamic interactions—showing, for example, how network limitations amplify profile‐sharing issues and shift children’s goal hierarchies in real time.

\subsubsection{Implications for design}
\begin{itemize}
  \item \textbf{Child-dedicated profiles.}  Implement automatic voice- or salutation-based profile switching to maintain separate histories, media preferences, and conversation contexts for children versus adults. This directly addresses prior observations of profile bleed on shared devices \cite{le_skillbot_2022} and preserves academic context when goals shift.
  \item \textbf{Data-sensitive interaction mode.}  Offer a “low-data” setting that downgrades multimedia responses to text and images under poor connectivity or on low-end hardware. This responds to children’s explicit “show me text” workarounds and ensures reliable service when bandwidth is constrained.
  \item \textbf{Parental oversight dashboard.}  Provide an intuitive interface for caregivers to define age-appropriate content filters, approve or blacklist sources, and allocate “homework vs.\ break” time blocks. This extends prior calls for better safety controls \cite{lovato_hey_2019} by aligning CA behaviour with family priorities and supporting joint parent–child regulation.
\end{itemize}

\subsection{RQ5: How can patterns in children’s CA interactions be translated into conversation-tree structures that better scaffold prompt formulation, problem solving, and engagement?}
Prior research on conversation design for child-CAs has shown that mimicking teacher-like dialogue structures—summarizing, questioning, and connecting to real-life examples—can boost engagement and comprehension \cite{garg_conversational_2020, xu_are_2021, xu_rosita_2023}. However, those studies largely prescribe static flows, without dynamically adapting to where a child is in their problem-solving process or supporting seamless multitasking. Our analysis advances this foundation by (a) identifying recurring scaffold patterns—direct help, indirect help, and hesitation—that naturally map onto decision nodes in a tree, and (b) weaving in QtA and ZPD strategies to diagnose a child’s state (confused vs. trial-and-error vs. reluctant) and offer tiered, context-sensitive hints. Unlike prior systems that treat each query in isolation, our conversation-tree model preserves context across academic, playful, and automation functions, supporting the multitasking rhythms children actually exhibit. In doing so, we both confirm the value of teacher-like questioning \cite{cagiltay_my_2023, ward_my_2011} and extend it by operationalizing these strategies in a branching structure that adapts in real time to a child’s successes, failures, and shifting goals. By weaving in Questioning the Author (QtA) \cite{ward_my_2011} and Zone of Proximal Development (ZPD) \cite{vygotsky_mind_1978} strategies, we can design baseline flows that (1) diagnose whether the child is confused, stuck in trial-and-error, or reluctant to ask for help, (2) offer tiered hints: from keyword suggestions to open-ended reflection prompts, and (3) prompt children to evaluate CA responses (“Does this answer make sense? What’s missing?”), thus fostering critical thinking.

\subsubsection{Implications for design}
\begin{itemize}
    \item \textbf{Model multitasking flows.} conversation-trees should enable transitions between academic queries, playful tangents, and task automation without “dropping” the prior context, thus supporting natural multitasking flows.
    \item \textbf{Hint mode. }Offer an MKO-inspired and QtA-style hint feature: when a child hesitates or repeats a prompt, the CA could interject with scaffolded suggestions (“Try asking: ‘List three examples of…’”; “Why do you think that answer fits your homework?”) to encourage reflection.
    \item \textbf{ZPD-tuned difficulty.} Adjust the specificity of hints based on the child’s history of successes and failures, ensuring challenges remain within reach. Prior work emphasizes reflection prompts \cite{ward_my_2011}, but without explicit checks, children may proceed without understanding. Checkpoints surface misunderstanding early and trigger appropriate scaffold branches.
    \item \textbf{Adaptive media suggestions.} At relevant nodes, proactively offer the most suitable media format—text, image, animation, or audio—based on the child’s past preferences and device capabilities. Prior studies show multimodal content boosts comprehension and engagement \cite{alaimi_pedagogical_2020, xu_rosita_2023}. Tailoring media in-flow ensures children receive information in their optimal learning modality without manual switching.
\end{itemize}

\subsection{RQ6: To what extent do conversation-tree recipes improve knowledge scaffolding across grade levels in CA interactions compared to unstructured, free-form dialogue in large-language models?}
While prior CA designs have largely relied on static scripts or narrow domain fine‑tunings that struggle to generalize across developmental stages and cultural contexts \cite{monteiro_investigating_2024, lee_dapie_2023}, our conversation‑tree recipe demonstrates a scalable path to dynamic, grade‑aligned scaffolding. Building on findings that children engage more deeply when CAs mimic teacher‑like questioning strategies—summarizing, probing for explanations, and connecting to real‑life examples \cite{garg_conversational_2020, xu_are_2021}—we structurally constrained a GPT 4o‑mini embedded with a conversation-tree recipe drawn from child‑parent scaffolding interactions. The result is an LLM constrained by a conversation-tree recipe (i.e.,structured prompt) that not only meets targeted readability levels across grade levels, but also generates an order of magnitude more scaffolded questions with  greater cognitive depth and interrogative diversity than vanilla LLM outputs. This contrasts  with rule‑based or single‑turn emotion‑tuned CAs \cite{garg_conversational_2020, xu_rosita_2023}, which often struggle to adjust question depth by grade level. By prescribing a conversation‑tree structure rather than fixed scripts, our recipe supports persistent context and dynamic turn‑management—features shown to enhance engagement in collectivist learning environments \cite{ogan_collaboration_2012}—while avoiding the brittleness of hard‑coded flows. In doing so, we offer a humanized CA framework that can adapt in real time to children’s developmental needs and cultural expectations, thereby advancing beyond prior one‑size‑fits‑all or monolithic LLM sampling strategies.

\subsection{Limitations and Future Work}
This study offers valuable insights into children's interactions with CAs, yet it has some limitations. The sample size of 23 participants, while informative, may not fully represent the diverse experiences of children in different cultural or socio-economic contexts, limiting the generalizability of the findings to other regions. However, we found in previous work that the problems Brazilian children encountered in our study were similar to those observed in other contexts \cite{kanda_two-month_2007, xu_elinors_2022, lovato_hey_2019, garg_conversational_2020}. Moreover, we applied the CWA \cite{rasmussen_taxonomy_1990} to ensure reliability and validity in our data analysis. Future work should include larger and more diverse samples to validate and expand upon our findings. Longitudinal studies could provide deeper insights into how children’s interactions with CAs evolve over time. Additionally, incorporating more advanced user experience metrics and exploring the impact of different CA functionalities across various cultural contexts could enhance the design recommendations and overall understanding of child-CA interactions.

Despite evidence that our conversation‑tree recipe improves grade‑aligned scaffolding and question prompting, several caveats apply. First, we used simulated child–CA exchanges rather than real classroom interactions, which may omit authentic peer collaboration and teacher cues. Future work should deploy the agent constrained by the strcutured prompt in‑situ—on low‑cost, offline devices in diverse schools—to evaluate engagement and learning gains. Moreover, our automated “question depth” and “diversity” metrics rely on interrogative words. Therefore, we will conduct a study with educators to human‑code question quality and refine these measures for stronger validity.

\section{Conclusion}
This study explored children's interactions with CAs in Brazilian settings, using Cognitive Work Analysis to understand their information processing and develop design recommendations. We found that children use CAs for Education, Discovery, Entertainment, and Task Automation, often concurrently and with support from family and peers.

Key design recommendations include creating four conversation-trees, child-dedicated profiles, and content curation. These recommendations are based on real-life strategies used by children, parents, teachers, and peers for problem-solving and decision-making.

Our contributions include cognitive work analysis for young populations in Latin America, insights into information processing flows in child-CA interactions, human-CA conversation-trees that blend human-human and human-computer communication, and integrated functions to support children in switching between educational, entertainment, and automation activities.

Although our research focuses on children in Latin America, these recommendations have broader applicability, guiding the development of CA dialogues and features that support effective prompt creation by blending human‑to‑human and human‑to‑computer interaction principles.

In summary, this study offers practical design recommendations that enhance CA design and child-CA interactions, supporting children's cognitive development and real-life engagement.
\begin{acks}
    We extend our heartfelt gratitude to the children, parents, and teachers who participated in this study. Your time, insights, and willingness to share your experiences with conversational agents were invaluable.
\end{acks}

\bibliographystyle{ACM-Reference-Format}
\bibliography{sample-base}

\end{document}